\documentclass[a4paper,11pt]{article}
\pdfoutput=1 

\usepackage{jheppub} 
\usepackage[T1]{fontenc} 
\usepackage{multirow}
\usepackage{rotating}
\usepackage{longtable}
\usepackage{dcolumn}  
\usepackage{lineno}
\include{patch_amsmath}
\usepackage{booktabs}
\usepackage{orcidlink}
\usepackage[tableposition=above]{caption}
\usepackage[shortlabels]{enumitem}

\renewcommand{\arraystretch}{1.1}

\newcommand{\mev}{\mathrm{MeV}}
\newcommand{\mevc}{\mathrm{MeV}/c}
\newcommand{\mevm}{\mathrm{MeV}/c^2}
\newcommand{\gev}{\mathrm{GeV}}

\newcommand{\gevm}{\mathrm{GeV}/c^2}

\newcommand{\ee}{e^+e^-}
\newcommand{\uu}{\mu^+\mu^-}

\newcommand{\Ufo}{\Upsilon(4S)}

\newcommand{\bp}{B^+}
\newcommand{\bm}{B^-}
\newcommand{\bn}{B^0}

\newcommand{\mbc}{M_{\rm bc}}

\newcommand{\DE}{\Delta E}
\newcommand{\fb}{\mathrm{fb}^{-1}}

\newcommand{\br}{\mathcal{B}}

\newcommand{\ecm}{E_{\mathrm{cm}}}
\newcommand{\ecmi}{E_{\mathrm{cm}}^{\mathrm{(i)}}}
\newcommand{\ez}{E_{\mathrm{cm0}}}

\newcommand{\spread}{\sigma_{\ecm}}
\newcommand{\deb}{\Delta E_{\mathrm{BaBar}}}
\newcommand{\nr}{n}
\newcommand{\sh}{s}
\newcommand{\ff}{\phi}

\newcommand{\bb}{B\bar{B}}
\newcommand{\bbst}{B\bar{B}^*}
\newcommand{\bstbst}{B^*\bar{B}^*}
\newcommand{\bball}{B^{(*)}\bar{B}^{(*)}}

\newcommand{\bnbn}{B^0\bar{B}^0}
\newcommand{\bpbm}{B^+B^-}
\newcommand{\dm}{\Delta m}
\newcommand{\bbbar}{b\bar{b}}
\newcommand{\ccbar}{c\bar{c}}
\newcommand{\dn}{D^0}
\newcommand{\rnp}{{\cal R}}
\newcommand{\rnpv}{{\cal R}_{\mathrm{v}}}
\newcommand{\rnpvi}{{\cal R}_{\mathrm{v}}^{\mathrm{(i)}}}
\newcommand{\mt}{\tilde{M}_{\rm bc}}
\newcommand{\aeb}{\bar{E}_B}
\newcommand{\ebb}{\bar{E}_{B\bar{B}}}

\newcommand{\aecm}{\bar{E}_{\rm cm}}
\newcommand{\dnx}{D^0/\bar{D}^0\,X}
\newcommand{\sigv}{\sigma_{\mathrm{v}}}

\newcommand{\NiH}{N^{\mathrm{(i)}}_{\operatorname{high-}x_p}}
\newcommand{\NiL}{N^{\mathrm{(i)}}_{\operatorname{low-}x_p}}
\newcommand{\rco}{r_{\rm cont}}

\def\beq{\begin{equation}}
\def\eeq{\end{equation}}
\def\bea{\begin{eqnarray}}
\def\eea{\end{eqnarray}}

\title{\boldmath Measurements of \\ the mass difference $m(B^0)-m(B^+)$ and \\ the energy dependence of the cross-section ratio $\sigma(\ee\to\bnbn)\,/\,\sigma(\ee\to\bpbm)$ \\at Belle and Belle~II}

\collaboration{The Belle and Belle II Collaborations}
  \author{M.~Abumusabh\,\orcidlink{0009-0004-1031-5425},} 
  \author{I.~Adachi\,\orcidlink{0000-0003-2287-0173},} 
  \author{L.~Aggarwal\,\orcidlink{0000-0002-0909-7537},} 
  \author{H.~Ahmed\,\orcidlink{0000-0003-3976-7498},} 
  \author{Y.~Ahn\,\orcidlink{0000-0001-6820-0576},} 
  \author{H.~Aihara\,\orcidlink{0000-0002-1907-5964},} 
  \author{N.~Akopov\,\orcidlink{0000-0002-4425-2096},} 
  \author{S.~Alghamdi\,\orcidlink{0000-0001-7609-112X},} 
  \author{M.~Alhakami\,\orcidlink{0000-0002-2234-8628},} 
  \author{A.~Aloisio\,\orcidlink{0000-0002-3883-6693},} 
% ...

\abstract{
Using data samples collected by the Belle and Belle~II experiments at
the $\Ufo$ resonance with integrated luminosities of $571\,\fb$ and
$365\,\fb$, respectively, we measure the pseudoscalar $B$-meson mass
difference to be $m(B^0)-m(B^+)=(0.495\pm0.024\pm0.005)\,\mevm$. The
results are based on a simultaneous fit to the variable $\mt$, which
is related to the $B$ momentum, for $\bn$ and $\bp$ candidates; and to
the energy dependence of
$\rnp=\sigma(\ee\to\bnbn)\,/\,\sigma(\ee\to\bpbm)$, which is measured
using changes in the average center-of-mass energy over the data
taking periods.
The phase-space hypothesis $\rnp=(p_{\bn}/p_{\bp})^3$, upon which
previous measurements rely, is strongly disfavored by our fit; the
measured mass-difference value for the phase-space hypothesis also
differs significantly from our measurement.
We constrain $\rnp$ in a broader energy range than covered by the
direct measurement and extract the energy dependence of $\rnp$ in the
range from the $\bb$ threshold up to $10.59\,\gev$. We interpret the
results using a phenomenological model and constrain the parameters of
the $\bb$ potential in the isovector channel.  }

\keywords{$B$ Physics, $e^+e^-$ Experiments, Quarkonium, Spectroscopy}

\begin{document}
\maketitle
\flushbottom

\section{Introduction}
\label{sec:intro}

The mass difference between the $\bn$ and $\bp$ mesons,
$\dm=m(\bn)-m(\bp)$, is an isospin-violating effect that arises due to
the mass difference between the $u$ and $d$ quarks and the
electromagnetic interaction between the $b$ quark and the light
antiquarks. The value of $\dm$ is sensitive to the masses of the $u$
and $d$ quarks and provides a constraint on the $B$-meson wave
function~\cite{Goity:2007fu}.
The energy dependence of the cross-section ratio
$\sigma(\ee\to\bnbn)\,/\,\sigma(\ee\to\bpbm)$ near the meson pair mass
threshold provides information about the $\bb$ potential in the
isovector channel~\cite{Dubynskiy:2007xw}, which is important for
understanding $\bball$ molecular states with isospin
one~\cite{Belle:2011aa}.

We measure $\dm$ using differences in the momentum distributions of
neutral and charged $B$-meson pairs produced at the $\Ufo$
resonance. The momentum distributions of $\bn$ and $\bp$ are different
for the following two reasons:
\begin{enumerate}[(a)]
\item The masses of $\bn$ and $\bp$ are different; this is the source
  of the sensitivity of our method.
\item The energy distributions of $\bn$ and $\bp$ are different due to
  the finite spread of the $\ee$ collision energy in the
  center-of-mass (c.m.) system, $\ecm$, and the different energy
  dependences of the $\ee\to\bnbn$ and $\ee\to\bpbm$ cross sections.
\end{enumerate}

For a precise measurement of $\dm$, it is essential to take correctly
into account the effect of (b), for which the energy dependence of the
cross-section ratio
\begin{equation}
  \rnp=\frac{\sigma(\ee\to\bnbn)}{\sigma(\ee\to\bpbm)}
\end{equation}
is required~\cite{Bondar:2022kxv}. However, the dependence
$\rnp(\ecm)$ has not been measured and there are no reliable
theoretical predictions for
it~\cite{Dubynskiy:2007xw,Voloshin:2003gm,Voloshin:2004nu}. Our
strategy is to measure both $\dm$ and $\rnp(\ecm)$ in this analysis;
to obtain $\rnp(\ecm)$ we use the variations of average $\ecm$ over
the Belle and Belle~II running periods. We use data from the Belle and
Belle~II experiments collected near the $\Ufo$ with integrated
luminosities of $571\,\fb$ and $365\,\fb$, respectively. (A $47\,\fb$
sample of Belle II data taken below the resonance, at $10.52\,\gev$,
is used to make an auxiliary measurement: see section~\ref{sec:d0x}.)

The current world-average of $\dm$ is dominated by the BaBar
measurement, $\dm=(0.33\pm0.05\pm0.03)\,\mevm$~\cite{BaBar:2008ikz},
which was determined using the average c.m.\ momenta of $\bn$ and
$\bp$ mesons produced at the $\Ufo$ resonance. For the energy
dependence of $\rnp$ it was assumed that
\begin{equation}
  \rnp = \left(p_{\bn} / p_{\bp} \right)^3.
  \label{eq:phsp}
\end{equation}
This assumption is not
justified~\cite{Dubynskiy:2007xw,Voloshin:2003gm,Voloshin:2004nu}; in
ref.~\cite{Bondar:2022kxv} it was pointed out that the change of
$\rnp$ with energy could be much faster than the expectation from
eq.~\eqref{eq:phsp}, which could lead to a bias in the $\dm$
measurement by as much as $0.4\,\mevm$.\footnote{ The RMS of the
  $\ecm$ spread is about $5\,\mev$, which corresponds to a FWHM of
  $12\,\mev$. This is only a factor of 0.6 lower than the width of the
  $\Ufo$. Thus, the energy interval spanned by the $\ecm$ spread is
  large, which explains why the effect of the $\rnp(\ecm)$ shape on
  the $\dm$ measurement is so significant.} Thus, it is important to
measure the energy dependence of $\rnp$.

To fully exploit the available information for the measurement of
$\dm$ and $\rnp(\ecm)$, we develop a combined fit to several
distributions that are sensitive to the quantities of interest. They
are the following (the details of each are described below in
sections~\ref{sec:mbc_fit}, \ref{sec:rat_vs_ecm}, \ref{sec:rb_fit},
and \ref{sec:d0x}): 
\begin{itemize}
\item The $\mt$ distributions of $\bn$ and $\bp$ candidates at Belle
  and Belle~II. The $\mt$ variable is uniquely related to the $B$
  momentum, 
\begin{equation}
  \mt=\sqrt{\left(\frac{m_{\Ufo}}{2}\right)^2-p_B^2},
  \label{eq:mt}
\end{equation}
where $m_{\Ufo}=10.58\,\gevm$ is approximately the known $\Ufo$
mass~\cite{ParticleDataGroup:2024cfk} and $B$ denotes $\bn$ or
$\bp$. The $\mt$ variable is more convenient than the momentum because
the shape of the combinatorial background near the kinematic boundary
$p_B=0$ is simpler. 

The $\mt$ distribution for $\bn$ ($\bp$) depends on the $\ecm$
distribution, the line-shape of the $\bnbn$ ($\bpbm$) cross section,
the $\bn$ ($\bp$) mass, and the momentum resolution of $\bn$
($\bp$)~\cite{Belle:2021lzm,Belle-II:2024niz}.
\item The energy dependence of $\rnp$ measured using variations in
  $\ecm$ over time. We measure the ratio of the $\bn$ and $\bp$ yields
  in slices of $\ecm$, which corresponds to $\rnp(\ecm)$.
\item The results of an energy scan for $\sigma(\ee\to\bbbar)$
  obtained by the BaBar experiment~\cite{Aubert:2008ab}. Below the
  $\bbst$ threshold,
  $\sigma(\ee\to\bbbar)=\sigma(\ee\to\bnbn)+\sigma(\ee\to\bpbm)$; the
  sum of the cross sections, together with $\rnp(\ecm)$, is needed to
  obtain the $\bn$ and $\bp$ line-shapes separately.
\item The cross section $\sigma(\ee\to\bbbar\to\dnx)$ versus
  $\ecm$. This constrains the energy scale of the
  $\sigma(\ee\to\bbbar)$ scan~\cite{Aubert:2008ab}.
\end{itemize}

The $\mt$ variable is similar to the widely used beam-constrained mass
\begin{equation}
  \mbc=\sqrt{\aeb^{\,2}-p_B^2},
  \label{eq:mbc}
\end{equation}
where $\aeb$ is the average energy of the $B$ meson for a given
running period, determined by calibration
(section~\ref{sec:rat_vs_ecm}). 
The $\mbc$ distributions for $\bn$ and $\bp$ candidates peak at the
corresponding nominal masses~\cite{ParticleDataGroup:2024cfk}. 
The resolution in $\mt$ is slightly worse than the resolution in
$\mbc$ (by a factor of 1.02 for Belle and by a factor of 1.09 for
Belle~II), since $\mt$ does not take into account the time variations
of the average $B$-meson energy. 
The variable $\mt$ is preferred since we can more accurately calculate
its shape in the fit function. 

The combined fit is sensitive to $\rnp(\ecm)$ in a wider energy range
than is covered by the time variations of the average collision
energy, because the $\mt$ distributions contain information about the
line-shapes of the $\bnbn$ and $\bpbm$ cross sections. The effective
sensitivity interval is the one spanned by the $\ecm$ spread. 
Therefore, we report results for $\rnp(\ecm)$ from the $\bb$ threshold
of $10.56\,\gev$ up to $10.59\,\gev$. We then perform a
phenomenological analysis of the $\rnp(\ecm)$ results to constrain
parameters of the isovector $\bb$ potential. 

Our work builds on the previous analyses of Belle~\cite{Belle:2021lzm}
and Belle~II \cite{Belle-II:2024niz}. We use the same samples of $\bn$
and $\bp$ mesons reconstructed in many hadronic final states, and we
use a $\mt$ fit function that takes into account the c.m.\ energy
spread, the energy dependence of the production cross section, and all
other relevant effects.  We also use a simultaneous fit to the $\mt$
distribution and measurements of the energy dependence of the
production cross section.  The novelty of the current analysis is the
use of the time variation of $\ecm$ for the cross-section
measurements.  Furthermore, in refs.~\cite{Belle:2021lzm} and
\cite{Belle-II:2024niz} the $\bn$ and $\bp$ mesons were combined; here
we consider them separately.

The paper is organized as follows. A brief description of the Belle
and Belle~II detectors and simulation is given in
section~\ref{sec:detector}. The selection of $\bn$ and $\bp$
candidates is summarized in section~\ref{sec:selection}. The function
used to fit the $\mt$ distributions is presented in
section~\ref{sec:mbc_fit}. Measurement of $\rnp$ using $\ecm$
variations is described in section~\ref{sec:rat_vs_ecm}. The fit to
the energy dependence of $\sigma(\ee\to\bbbar)$ and the measurement of
$\sigma(\ee\to\bbbar\to\dn\,X)$ are presented in
sections~\ref{sec:rb_fit} and \ref{sec:d0x}, respectively. The results
of the combined fit are given in section~\ref{sec:results}. Systematic
uncertainties and additional studies are described in
section~\ref{sec:syst}. The measurement of $\rnp$ in a wide energy
range and its phenomenological analysis are reported in
sections~\ref{sec:r_vs_ecm_wide} and \ref{sec:pheno},
respectively. Finally, the conclusions are given in
section~\ref{sec:conc}.

\section{Detectors and simulation}
\label{sec:detector}

The Belle experiment~\cite{Belle:2000cnh,Belle:2012iwr} operated at
the KEKB asymmetric-energy $\ee$
collider~\cite{Kurokawa:2001nw,Abe:2013kxa} between 1999 and
2010. KEKB used 8\,GeV electrons and 3.5\,GeV positrons. The detector
consisted of a large-solid-angle spectrometer, which included a
double-sided silicon-strip vertex detector, a 50-layer central drift
chamber, an array of aerogel threshold Cherenkov counters, a
barrel-like arrangement of time-of-flight scintillation counters, and
an electromagnetic calorimeter composed of CsI(Tl) crystals. All
subdetectors were located inside a superconducting solenoid coil that
provided a 1.5~T magnetic field. An iron flux-return yoke, placed
outside the coil, was instrumented with resistive-plate chambers to
detect $K^0_L$ mesons and identify muons. Two inner detector
configurations were used: a 2.0\,cm radius beam pipe and a three-layer
silicon vertex detector; and, from October 2003, a 1.5\,cm radius beam
pipe, a four-layer silicon vertex detector, and a small-inner-cell
drift chamber~\cite{Natkaniec:2006rv}. In this paper, we use data from
this second inner detector configuration only.

The Belle~II detector~\cite{Belle-II:2010dht,Belle-II:2018jsg} is an
upgrade with several new subdetectors designed to handle the
significantly larger beam-related backgrounds of the new SuperKEKB
$e^+e^-$ collider~\cite{Akai:2018mbz}. SuperKEKB uses 7\,GeV electrons
and 4\,GeV positrons, which results in the same c.m.\ energy as at
KEKB.
The detector consists of a silicon vertex detector wrapped around a
1.0\,cm radius beam pipe that contains two inner layers of pixel
detectors and four outer layers of double-sided strip detectors, a
56-layer central drift chamber, a time-of-propagation detector, and an
aerogel Cherenkov detector. The Belle CsI(Tl) crystal calorimeter, the
Belle solenoid, and the iron flux-return yoke are reused in the Belle
II detector. The calorimeter readout electronics have been upgraded,
and the endcaps of the flux-return yoke, together with the two
innermost layers of the barrel, have been re-instrumented with plastic
scintillator modules; the read-out system has also been upgraded. For
the data used in this paper, collected between 2019 and 2022, only
part of the second layer of the pixel detector, covering 15\% of the
azimuthal angle, was installed.

The data analysis strategy is tested on simulated event samples.
We generate $\ee\to\Ufo\to\bb$ events and simulate particle decays
with EvtGen, interfaced to Pythia~\cite{Lange:2001uf}; we generate
continuum $\ee\to{}q\bar{q}$ (where $q$ is a $u$, $d$, $c$, or $s$
quark) with Pythia6~\cite{Sjostrand:2006za} for Belle, and with
KKMC~\cite{Ward:2002qq} and Pythia8~\cite{Sjostrand:2014zea} for
Belle~II; we simulate detector response using
Geant3~\cite{Brun:1994aa} for Belle and Geant4~\cite{GEANT4:2002zbu}
for Belle~II.

\section{Event selection}
\label{sec:selection}

We use the Belle~II analysis software framework (basf2) to reconstruct
both Belle and Belle~II data~\cite{Kuhr:2018lps,basf2}. The Belle data
are converted to the Belle II format for basf2 compatibility using the
B2BII framework~\cite{Gelb:2018agf}.

We fully reconstruct $\bn$ and $\bp$ mesons in about 1000 exclusive
hadronic final states using the Full Event Interpretation (FEI)
package~\cite{Keck:2018lcd}. Background suppression in this package is
based on a multivariate analysis. We use a version of the FEI adapted
for energy-dependent cross-section measurements, described in
refs.~\cite{Belle:2021lzm,Belle-II:2024niz}.

\begin{figure}[htbp]
  \centering
  \setlength{\unitlength}{0.1\textwidth}
  \includegraphics[width=0.48\linewidth]{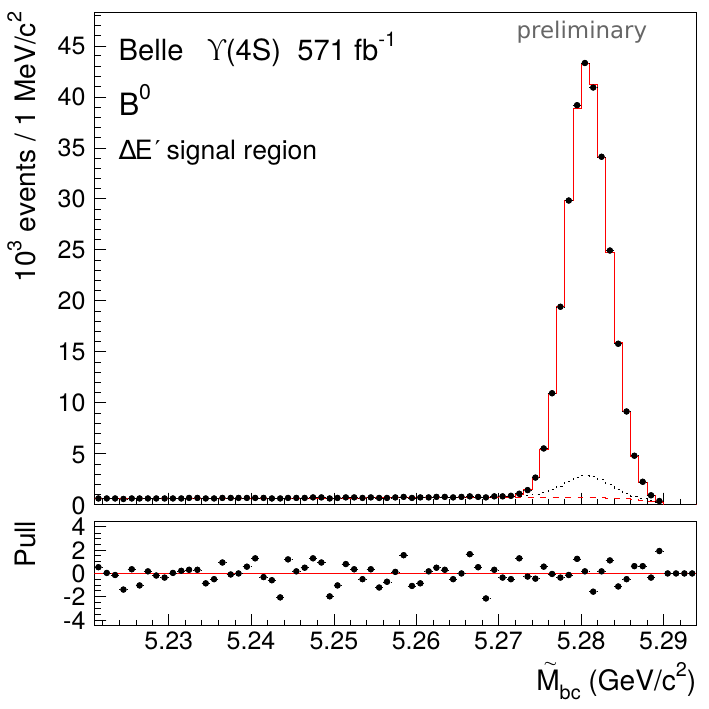}\hfill
  \includegraphics[width=0.48\linewidth]{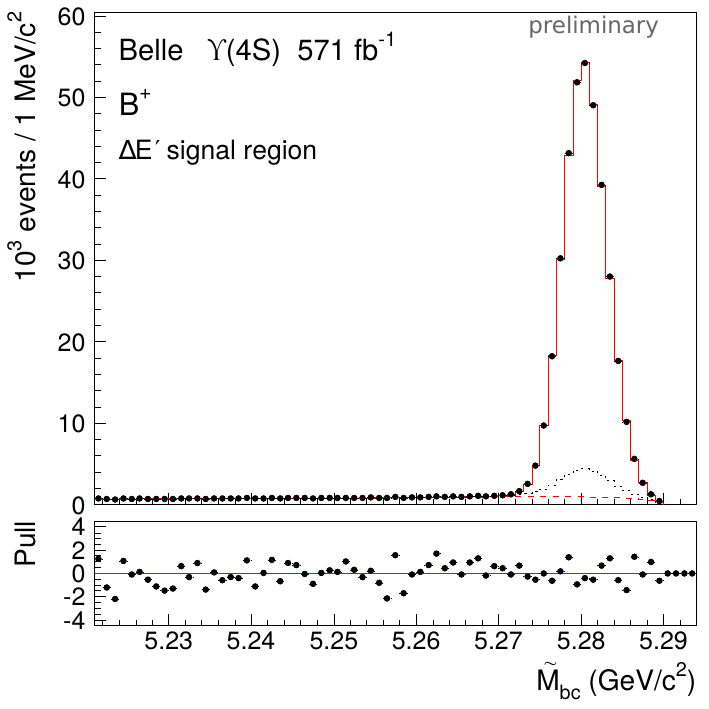}\vspace{1mm}
  \includegraphics[width=0.48\linewidth]{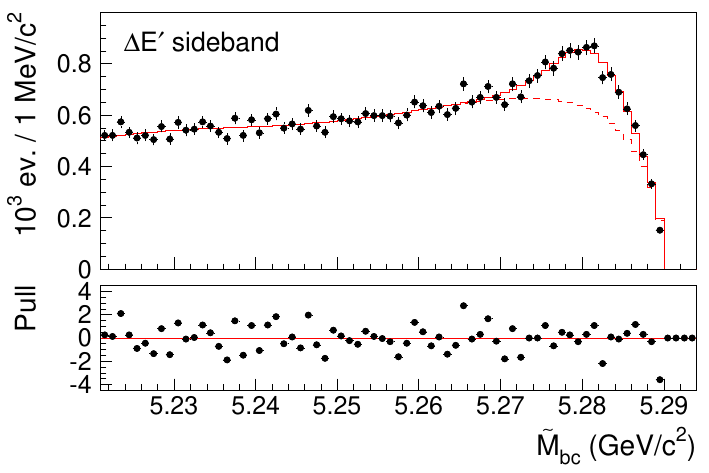}\hfill
  \includegraphics[width=0.48\linewidth]{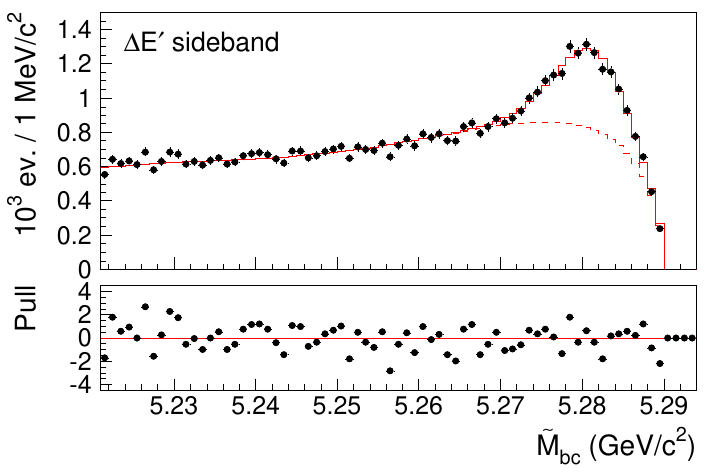}
  \caption{The $\mt$ distributions in the Belle $\Ufo$ data for the
    $\bn$ (left) and $\bp$ (right) candidates. The top and bottom
    parts of the figure correspond to the $\DE'$ signal region and
    sideband, respectively. The red solid histogram is the result of
    the combined fit described in section~\ref{sec:results}. The red
    dashed histogram shows the background, and the black dotted
    histogram shows the sum of the background and the broken signal
    (for the $\DE'$ sideband, this coincides with the total fit, and
    is thus not visible). The bottom panels show pulls (deviations of
    the data points from the fit function divided by the uncertainties
    on the data). }
  \label{mbc_y4s_bp_belle_280324}
\end{figure}

\begin{figure}[htbp]
  \centering
  \setlength{\unitlength}{0.1\textwidth}
  \includegraphics[width=0.48\linewidth]{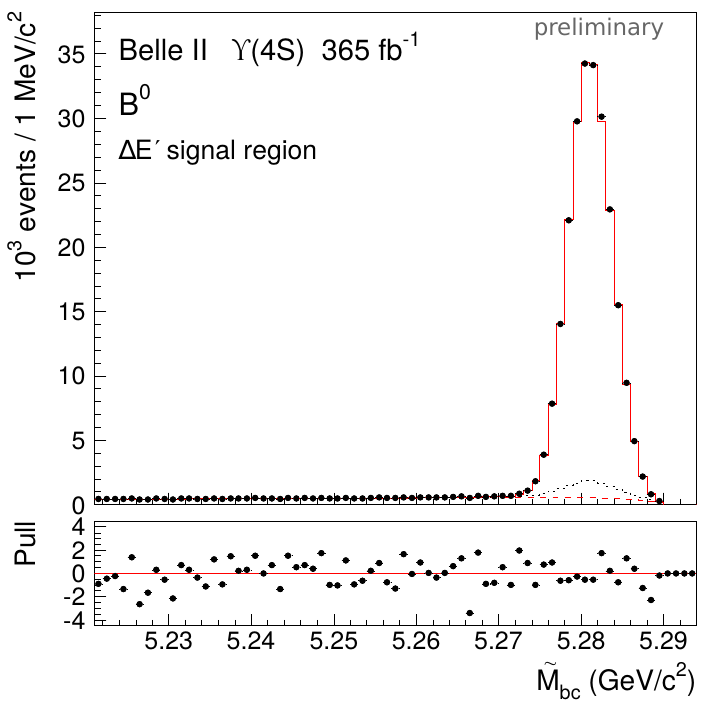}\hfill
  \includegraphics[width=0.48\linewidth]{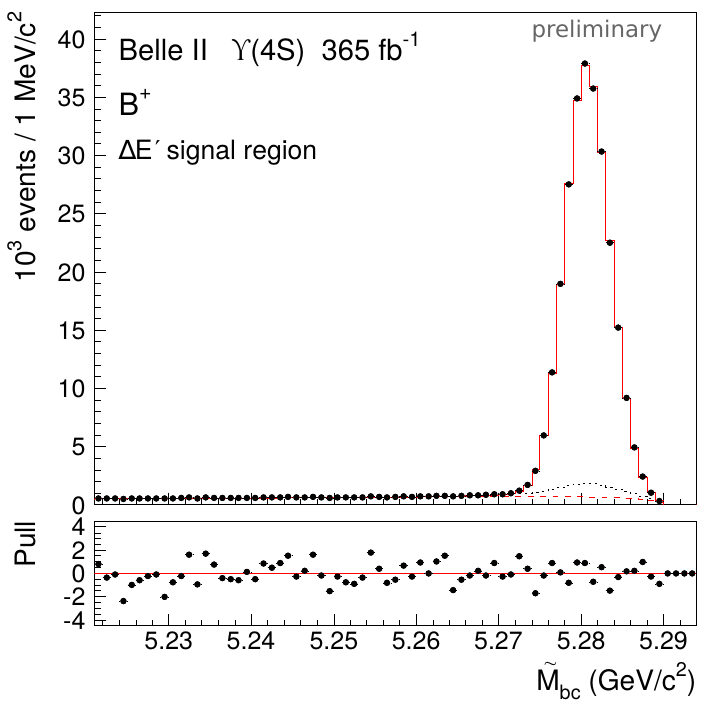}\vspace{1mm}
  \includegraphics[width=0.48\linewidth]{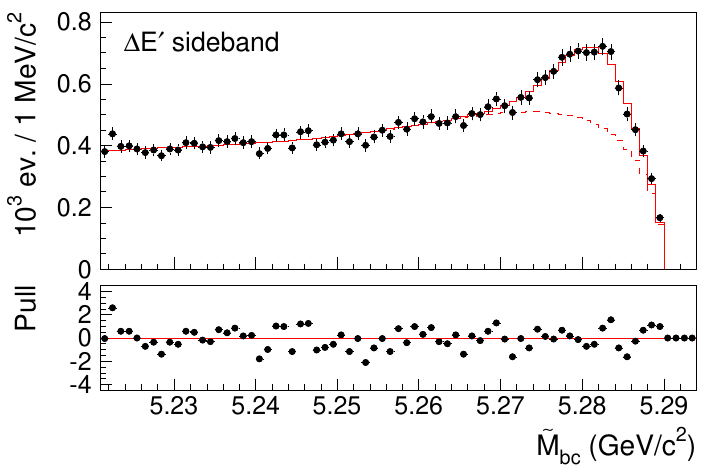}\hfill
  \includegraphics[width=0.48\linewidth]{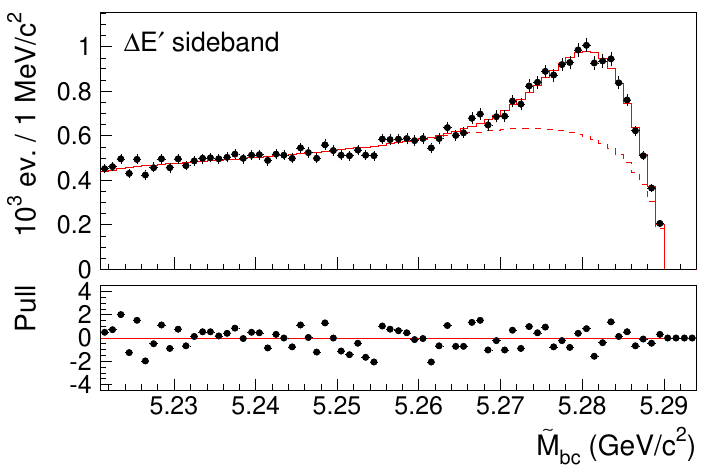}
  \caption{ The $\mt$ distributions in the Belle~II $\Ufo$ data for
    the $\bn$ (left) and $\bp$ (right) candidates. The panels and the
    curves have the same meaning as in
    figure~\ref{mbc_y4s_bp_belle_280324}. }
  \label{mbc_y4s_bp_belle2_280324}
\end{figure}

Signal candidates peak at zero in the $\DE$ distribution,
\begin{equation}
  \DE = E_B - \aeb,
\end{equation}
where $E_B$ is the energy of the $B$ candidate in the $\ee$
c.m.\ system.
The sum $\DE+\mbc$ is equal to the mass of the $B$ candidate to a high
precision. Therefore, in the variable
\begin{equation}
  \DE' = \DE + \mbc - m_B,
\end{equation}
the effect of the energy spread cancels.
We select $B$ candidates in the $\DE'$ signal region and sideband. In
the case of Belle~II, the $\DE'$ signal region is defined as
$|\DE'|<18\,\mev$, while in the case of Belle, the size of the signal
region varies depending on the reconstruction channel. 
The efficiency of the $\DE'$ requirement is close to 92\% for both Belle and Belle~II.
The sideband is shifted by $+80\,\mev$ and has the same width as the
signal region and is used to constrain the background shape.
The $\mt$ distributions for selected $\bp$ and $\bn$ candidates in the
$\DE'$ signal region and sideband in the Belle and Belle~II data
samples are shown in figures~\ref{mbc_y4s_bp_belle_280324} and
\ref{mbc_y4s_bp_belle2_280324}.

\section{\boldmath $\mt$ fit function}
\label{sec:mbc_fit}

The $\mt$ signal fit function is a key tool in this analysis. It is
based on the fit functions developed in
refs.~\cite{Belle:2021lzm,Belle-II:2024niz}. The fit function is
calculated numerically at each minimization step and takes into
account the following: 
\begin{itemize}
\item variation of the average $\ee$ c.m.\ energy $\aecm$ over the
  data-taking period;\footnote{The change of collision energy with
    time was not explicitly considered in
    refs.~\cite{Belle:2021lzm,Belle-II:2024niz}; thus, the
    corresponding effect was absorbed in the energy spread.} 
\item $\ecm$ spread due to fluctuations in synchrotron radiation emission, $\spread$;
\item initial state radiation (ISR);
\item energy dependence of the $\ee\to\bnbn$ and $\ee\to\bpbm$ cross sections;
\item momentum resolution of reconstructed $B$ mesons.
\end{itemize}
The calculation scheme for the $\bp$ fit function is as follows:
\begin{gather}
  [\sum_i w_{\mathrm{i}}\;G(\ecm; \;\ez+\Delta\ecmi,\,\spread)] \otimes f_{\mathrm{ISR}} \times \sigma(\ee\to\bpbm) \nonumber\\
   \longmapsto f(p_{\bp}),\, \otimes f_{\mathrm{resolution}} \longmapsto f(\mt), 
\label{eq:scheme}
\end{gather}
where $G(\ecm)$ is a Gaussian representing the c.m.\ energy spread;
$\ez$ is the c.m.\ energy averaged over the entire running period;
$\Delta\ecmi$ is the energy deviation of the $i$-th data subsample;
$w_i$ is the corresponding weight (the fraction of the total sample in
the $i$-th subsample); $\sum_i$ is the sum over different subsamples;
and $\otimes f_{\mathrm{ISR}}$ represents convolution with the ISR
kernel~\cite{Kuraev:1985hb}. At this stage of the calculation, we
obtain the distribution in the invariant mass of the virtual photon
produced in $\ee$ annihilation. Next, we multiply this distribution by
the energy dependence of the $\bpbm$ cross section [$\times
  \sigma(\ee\to\bpbm)$], obtaining the distribution of the energy of
$\bpbm$ pairs. We then change variables from $\bpbm$ energy to $\bp$
momentum (denoted by ``$\longmapsto$''). This change takes into
account nonlinear effects, which are especially important near the
kinematic boundary $p_{\bp}=0$. At this stage, the value of $m(\bp)$
enters the fit function. 
We perform a convolution to account for the $\bp$ momentum resolution
(``$\otimes f_{\mathrm{resolution}}$''). Finally, we change variables
from $p_{\bp}$ to $\mt$. 
The fit function for $\bn$ is calculated in a similar way, but using
the $\bnbn$ cross section, the $\bn$ mass, and the $\bn$
momentum-resolution. 

The parameters $\ez$ and $\spread$ are determined from the combined
fit. The deviations $\Delta\ecmi$ and the weights $w_i$ are determined
using the $\mt$ fits in $\aeb$ bins (section~\ref{sec:rat_vs_ecm}
below).  
There are separate parameters for the Belle and Belle~II data samples.
The energy dependence of the dressed cross sections\footnote{The
  difference between the Born and dressed cross sections is that the
  latter includes the vacuum polarization effect.} is parameterized
as~\cite{Belle-II:2024niz} 
\begin{align}
  \sigma(\ee\to\bpbm) & = p_{\bp}^3\;P_{11}(\ecm), \label{eq:ee_to_bpbm} \\
  \sigma(\ee\to\bnbn) & = p_{\bn}^3\;P_{11}(\ecm)\;P_2(\ecm), \label{eq:ee_to_bnbn}
\end{align}
where $p_{\bn}^3$ and $p_{\bp}^3$ are phase-space factors,
$P_{11}(\ecm)$ is an 11th order Chebyshev polynomial which is common
for both cross sections, and $P_2(\ecm)$ is a second order Chebyshev
polynomial that accounts for additional energy dependence of the
cross-section ratio beyond the phase-space factor. All parameters of
the Chebyshev polynomials are allowed to float in the combined fit. 
The mass of $\bp$ is fixed to the average value of the Particle Data
Group (PDG),
$m(\bp)=(5279.42\pm0.08)\,\mevm$~\cite{ParticleDataGroup:2024cfk},
which, unlike the PDG fit value, does not include the $\dm$
measurement of BaBar~\cite{BaBar:2008ikz}. 
The $\bn$ and $\bp$ mass difference $\dm$ is determined from the
combined fit. Alternatively, one could fix $m(\bn)$ and vary $\dm$; we
chose to fix $m(\bp)$ because it is 2.5 times more precise.

The $B$-meson momentum resolution functions are obtained from the
simulation and are adjusted to take into account the difference
between simulation and data. 
In the simulation, we find three types of events that peak in $\mt$
distributions: (1) correctly reconstructed signal candidates, (2)
broken-signal candidates in the $\DE'$ signal region, and (3)
broken-signal candidates in the $\DE'$ sideband. The broken signal is
due to signal events in which one of the final-state particles
originates from background. 
The momentum resolution for the correctly reconstructed candidates is
described by a sum of three Gaussian functions multiplied by a factor
taking into account the fact that momentum is required to be
positive~\cite{Belle-II:2024niz}. 
The broken signal components are described via a similar function but
with a large width~\cite{Belle:2021lzm,Belle-II:2024niz}. 
The yield ratios of the broken and correctly reconstructed signal
candidates are about 5\% and 2\% for the $\DE'$ signal region and
sideband, respectively. The yield ratios and parameters of the
momentum resolution functions are fixed using simulation (the values
are provided in Refs.~\cite{Belle:2021lzm,Belle-II:2024niz}). 
To account for the difference between simulation and data, we
introduce shifts, $\sh_i$, and broadening factors, $\ff_i$, where $i$
runs over all event types, $i=1,2,3$. For broken signal, in addition
we introduce normalization corrections, $\nr_2$ and $\nr_3$. There are
separate corrections for $\bn$ and $\bp$, and for the Belle and
Belle~II data samples. 
The broadening factors for correctly reconstructed candidates,
$\ff_1$, are estimated using the fits to the $\DE'$ distributions in
the $\mt$ signal region~\cite{Belle:2021lzm,Belle-II:2024niz}. The
broadening factor of $\DE'$ is $1.18\pm0.01$ for Belle and
$1.07\pm0.01$ for Belle~II. From the simulation, we find that the
broadening factor in $\DE'$ is larger than that in the momentum
resolution by 3\%. We account for this correction and fix $\ff_1$ in
the combined fit. We also fix $\sh_1=\sh_2=0$ and $\nr_2=\ff_2=1$,
while the parameters $\nr_3$, $\sh_3$ and $\ff_3$ are determined from
the fit. 

Combinatorial background is parameterized by the threshold function,
$\sqrt{\frac{m_{\Ufo}}{2}-\mt}$, multiplied by a second order
polynomial. The shape of the combinatorial background is the same in
the $\DE'$ signal region and sideband, while the normalizations
differ. All parameters of the combinatorial background are determined
from the fit.

\section{\boldmath Measurement of the energy dependence of $\rnp$}
\label{sec:rat_vs_ecm}

\begin{figure}[htbp]
\centering
\includegraphics[width=0.45\linewidth]{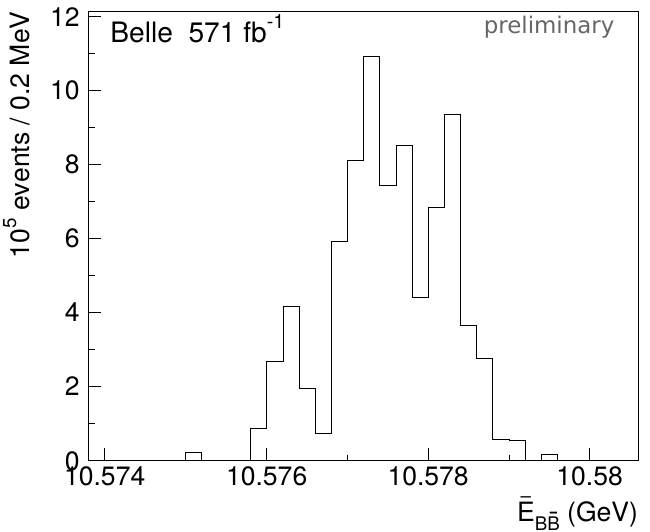}\hspace{7mm}
\includegraphics[width=0.45\linewidth]{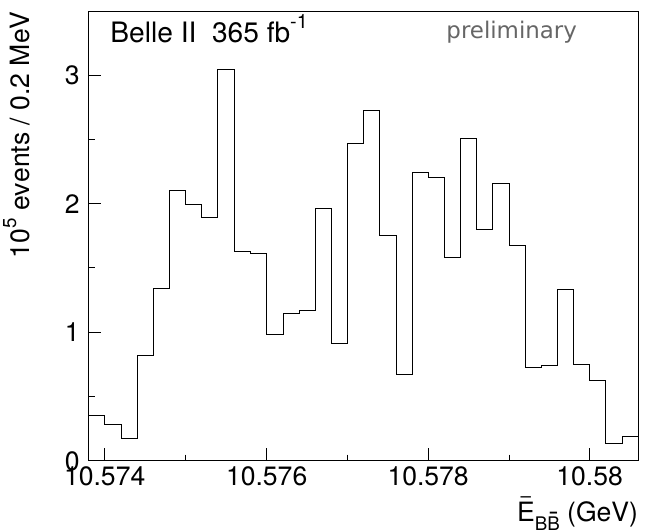}
\caption{ The distribution of $B$-meson candidates in the calibrated
  energy of the $\bb$ pair in Belle (left) and Belle~II data (right).}
\label{ecm_belle_270225}
\end{figure}

\begin{figure}[htbp]
\centering
\includegraphics[width=0.51\linewidth]{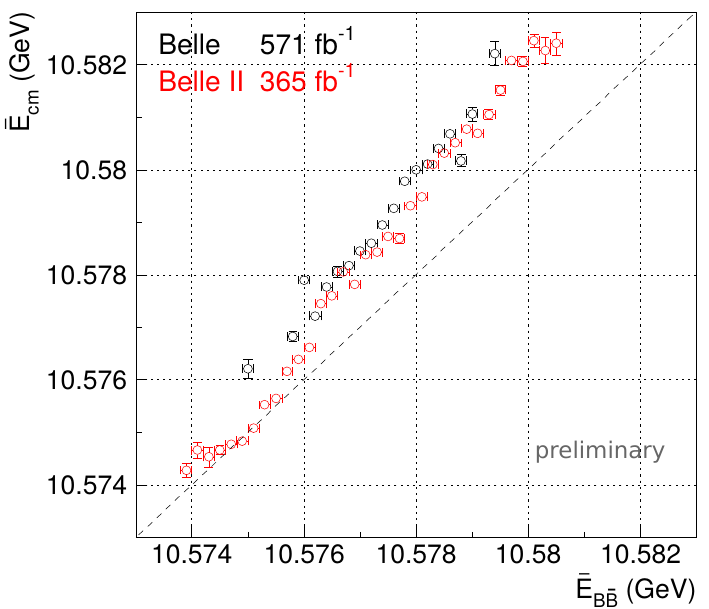}
\caption{ Energy of the colliding beams as a function of the energy of
  the $\bb$ pair. The black and red dots correspond to the Belle and
  Belle~II data, respectively. Horizontal error bars indicate $\ebb$
  bins, while vertical error bars show the statistical uncertainty in
  $\aecm$.
}
\label{ecm_vs_ebb_260224}
\end{figure}

The $\rnp(\ecm)$ dependence is measured using variations in the
collision energy over the course of data taking. 
The $B$-meson energy is calibrated in the following
way~\cite{BelleII_beam}: the total data sample is divided into parts
corresponding to data collection periods of several days. The average
energy $\aeb$ is then determined using several low-multiplicity,
low-background $B$-decay channels. The $\aeb$ accuracy achieved for
one part is typically $0.1-0.2\,\mev$.\footnote{The calibration of the
  boost vector, i.e.\ the velocity of the c.m.\ system in the
  laboratory frame, is based on the angles of the muons in the
  $\ee\to\uu$ process~\cite{BelleII_beam}.}
The distribution of the selected $B$-meson candidates in the energy of
the $\bb$ pair, $\ebb\equiv2\,\aeb$, is shown in
figure~\ref{ecm_belle_270225}. 
The $\ebb$ values span intervals of width $4.4\,\mev$ at Belle and
$6.6\,\mev$ at Belle~II. 
We subdivide the $\ebb$ range into bins of width $0.2\,\mev$ and
perform a simultaneous fit to the $\mt$ distributions of the $\bn$ and
$\bp$ candidates in each bin.  In the $i$-th bin, we fix all
parameters of the $\mt$ fit function except the total $\bn$ and $\bp$
signal yield, $(N_{\bn}+N_{\bp})^{\mathrm{(i)}}$; the yield ratio,
$(N_{\bn}/N_{\bp})^{\mathrm{(i)}}$; the average c.m.\ energy
$\aecm^{\mathrm{(i)}}$ for a given $\ebb$ bin; the $\ecm$ spread; and
the combinatorial background yields.
The results for $\aecm$ as a function of $\ebb$ are shown in
figure~\ref{ecm_vs_ebb_260224}. 
The values of $\aecm$ span the intervals of $5.9\,\mev$ and
$8.1\,\mev$ at Belle and Belle~II, respectively. These intervals are
wider than those of $\ebb$ by a factor of $\sim1.3$. This behavior
agrees with expectations based on
eq.~\eqref{eq:scheme}.\footnote{Indeed, in the absence of the ISR, the
  $\ebb$ value is always shifted from $\aecm$ towards the
  cross-section peak (the shift is zero if $\aecm=m_{\Ufo}$), which
  shrinks the interval spanned by $\ebb$ compared to that spanned by
  $\aecm$. The effect of ISR is to further shift all the points
  towards lower $\ebb$ values.} 
The results for the energy spread are consistent with being
independent of $\ebb$ (the absolute values for Belle and Belle~II are
given in section~\ref{sec:results}). 

The ratio of visible $\ee\to\bnbn$ and $\ee\to\bpbm$ cross sections is
determined as 
\begin{equation}
  \rnpvi=(N_{\bn}/N_{\bp})^{\mathrm{(i)}}\,(1/r_{\varepsilon}),
\end{equation}
where $r_{\varepsilon}\equiv\varepsilon_{\bn}/\varepsilon_{\bp}$ is
the ratio of the $\bn$ and $\bp$ reconstruction efficiency, the change
of which with energy is negligibly small. 
We do not attempt to determine $r_{\varepsilon}$ from simulation as
the corresponding systematic uncertainty could be difficult to
control. Instead, for the absolute normalization of $\rnpv$, we use
the world-average value of the measurements at the energy of the
$\Ufo$ peak~\cite{HeavyFlavorAveragingGroupHFLAV:2024ctg} 
\begin{equation}
  \rnpv=0.951\pm0.028.
  \label{eq:abs_fnp}
\end{equation}
We apply the constraint of eq.~\eqref{eq:abs_fnp} in the combined
fit. The error in eq.~\eqref{eq:abs_fnp} is accounted for as a
systematic uncertainty (section~\ref{sec:syst}). The values of
$r_{\varepsilon}$ at Belle and Belle~II are determined from the
combined fit. 

\begin{figure}[htbp]
\centering
\includegraphics[width=0.48\linewidth]{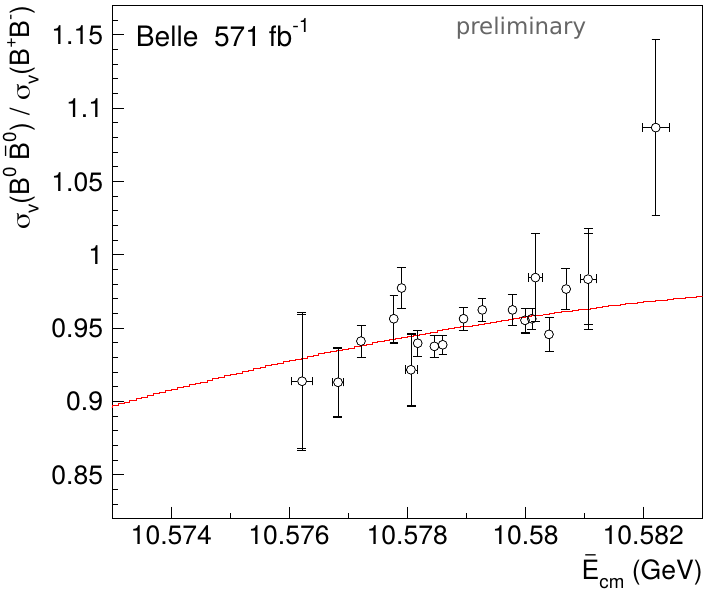}\hfill
\includegraphics[width=0.48\linewidth]{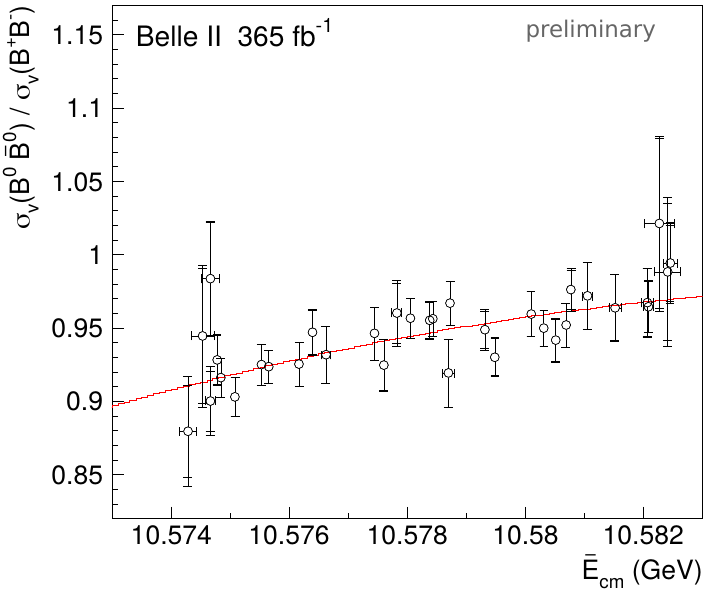}
\caption{ Energy dependence of the ratio of visible cross sections
  $\sigv(\ee\to\bnbn)\,/\,\sigv(\ee\to\bpbm)$. The dots with error
  bars show the direct measurements performed using Belle (left) and
  Belle~II (right) data; the vertical inner and outer error bars show
  statistical and total uncorrelated uncertainties, respectively; the
  horizontal error bars show statistical uncertainty. The curve is the
  result of the combined fit described in section~\ref{sec:results}. }
\label{rat_vs_ecm_small_280324}
\end{figure}

The ratio of visible $\ee\to\bnbn$ and $\ee\to\bpbm$ cross sections as
a function of $\aecm$ is shown in
figure~\ref{rat_vs_ecm_small_280324}. 
The Belle and Belle~II results are in good agreement; there is a
slight increase of $\rnpv$ with energy. 
The fit function describing the $\rnpv$ measurements is the ratio of
visible $\sigv(\ee\to\bnbn)$ and $\sigv(\ee\to\bpbm)$ cross sections
that are obtained from dressed cross sections in
Eqs.~\eqref{eq:ee_to_bpbm} and \eqref{eq:ee_to_bnbn} by applying
radiative corrections (convolving with the radiative kernel) and
accounting for the beam-energy spread (convolving with a Gaussian) as
described in refs.~\cite{Belle:2021lzm,Belle-II:2024niz}. 

In the $\mt$ fits in energy bins, many parameters are fixed to the
result of the combined fit. To obtain self-consistent results, we use
an iterative procedure. We perform the combined fit, fix parameters
and perform $\mt$ fits in $\aeb$ bins to measure $\rnpv$. The latter
results are used in the next iteration. The fit converges after two
iterations. 

To estimate the systematic uncertainties in $\rnpvi$ due to
uncertainties in the parameters that are fixed in the $\mt$ fits, we
use a pseudo-experiment technique. We generate pseudo-experiments
using the fitted function that is obtained from the combined fit
(section~\ref{sec:results}). We then perform a combined fit to the
pseudo-experiments and use the results to fix the parameters of the
$\mt$ fits in $\ebb$ bins. We repeat the measurements of $\rnpv$ in
$\ebb$ bins using parameters of each pseudo-experiment. We treat the
RMS of the deviations of $\rnpvi$ as a systematic uncertainty, which
is found to be negligibly small. 

To estimate the systematic uncertainty due to the modelling of the
smooth background, we float the parameters of the corresponding
polynomial in the fits in the $\ebb$ bins. Deviations of $\rnpvi$ are
treated as symmetric systematic uncertainties. These uncertainties are
small compared to statistical uncertainties and are assumed to be
uncorrelated for different $\ebb$ bins. The total uncertainty is
obtained by summing the statistical and systematic contributions in
quadrature. The total and statistical uncertainties are shown in
figure~\ref{rat_vs_ecm_small_280324}.

\section{\boldmath $R_b$ fit function}
\label{sec:rb_fit}

\begin{figure}[htbp]
\centering
\includegraphics[width=0.65\linewidth]{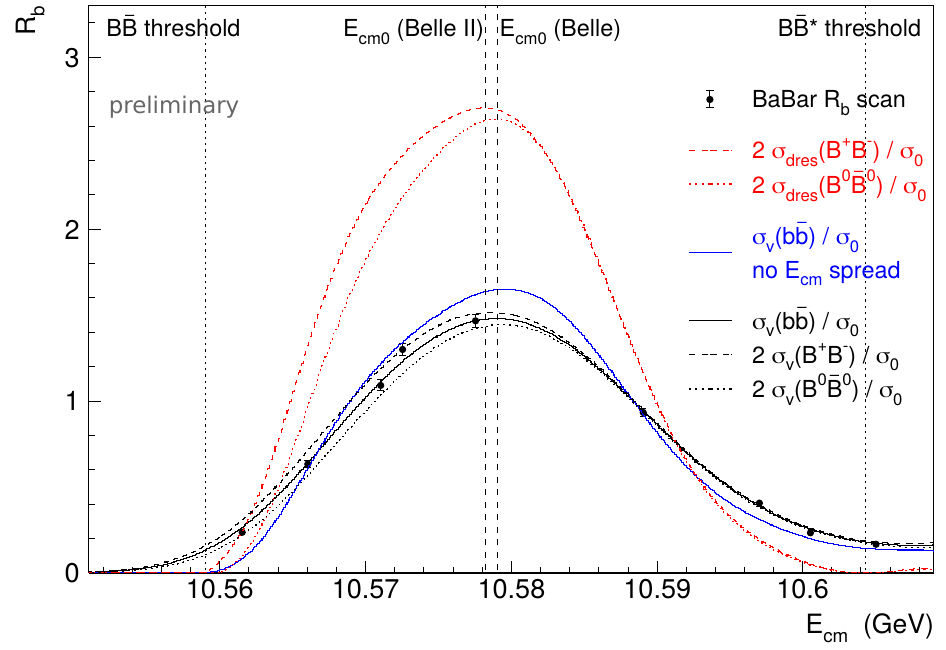}
\caption{ Energy dependence of $R_\mathrm{b}$. 
  The points with error bars are measurements from BaBar~\cite{Aubert:2008ab}.
  The black solid curve is the result of the combined fit described
  in section~\ref{sec:results}; the black dashed and dotted curves are
  the contributions of $\bp$ and $\bn$, respectively, multiplied by a
  factor of two; the blue solid curve corresponds to the visible cross
  section before accounting for the $\ecm$ spread; the red dashed and
  dotted curves correspond to the dressed cross sections of the $\bp$
  and $\bn$ production, respectively, multiplied by a factor of
  two. The vertical dashed lines show the average over the data taking
  period c.m.\ energies of Belle (right) and Belle~II (left), the
  vertical dotted lines indicate the $B\bar{B}$ and the $B\bar{B}^*$
  thresholds. }
\label{xsec_y4s_280324}
\end{figure}

BaBar reported the most precise measurement of the total visible
$\ee\to\bbbar$ cross section in the region of the
$\Ufo$~\cite{Aubert:2008ab}. The results were presented in terms of
$R_b$,
\begin{equation} 
  R_b=\frac{\sigv(\ee\to\bbbar)}{\sigma_0},
\end{equation}
where $\sigma_0=\frac{4\pi\alpha}{3}\,\frac{1}{\ecm^2}$ is the
Born-level $\ee\to\uu$ cross section, and $\alpha$ is the QED
fine-structure constant. The results of the BaBar scan for $R_b$ are
shown in figure~\ref{xsec_y4s_280324}. 
As in the previous publications~\cite{Belle:2021lzm,Belle-II:2024niz},
we include the BaBar measurements of $R_b$ at nine energies between
the $\bb$ and $\bbst$ thresholds into the combined fit. 
The fit function describing the energy dependence of $R_b$ is
$[\sigv(\ee\to\bnbn)+\sigv(\ee\to\bpbm)]/\sigma_0$. 
The BaBar scan points have a rather large energy scale uncertainty of
$1.5\,\mev$~\cite{BaBar:2008ikz}. We allow all the $R_b$ points to be
shifted simultaneously along the horizontal axis by a parameter
$\deb$, which is included in the fit with a Gaussian constraint around
zero and a width of $\sigma=1.5\,\mev$.

\section{\boldmath Measurement of the energy dependence of $\sigma(\ee\to\bbbar\to \dnx)$}
\label{sec:d0x}

In this analysis, the combined fit includes the energy dependence of
the inclusive $\ee\to\bbbar\to\dn\,X$ cross section. This helps to
constrain the peak position of the $R_b$ scan~\cite{Aubert:2008ab},
which otherwise has a large uncertainty as noted in the previous
section. We measure the energy dependence of the cross section using,
as in the case of $\rnp$, variations of $\ecm$ over the data-taking
period. We use an approach in which the luminosity of data samples at
various energies is determined using the yield of
$\ee\to\ccbar\to\dn\,X$. In this approach, the uncertainties due to
changes in the detector performance over the data taking period
largely cancel. 

To measure the $\ee\to\bbbar\to\dnx$ cross section, we reconstruct
$\dn\to{}K^-\pi^+$ decays in the $\Ufo$ data of Belle~II. The $K^-$
and $\pi^+$ candidates must originate from the interaction region and
be positively identified based on the energy-loss measurements in the
drift chamber and the response of the particle identification
systems. The selection requirements are the same as in
ref.~\cite{Belle:2023yfw}. 

For each $\ebb$ bin, we determine the $\dn$ yields in the $x_p<0.5$
and $x_p>0.55$ regions, $\NiL$ and $\NiH$. The normalized momentum
$x_p$ is defined as $x_p=p_D/p_D^{\mathrm{max}}$, where $p_D$ is the
c.m.\ momentum of the $\dn$ candidate, $p_D^{\mathrm{max}}$ is the
maximal kinematically allowed value of $p_D$,
$p_D^{\mathrm{max}}=\sqrt{(\aecm/2)^2-m_D^2}$, and $m_D$ is the
$\dn$-meson world-average mass~\cite{ParticleDataGroup:2024cfk}. 
For $\dn$ mesons produced in $\ee\to\bbbar$ events, $x_p$ is below
0.55~\cite{Belle:2023yfw}. In the case of continuum $\ee\to\ccbar$,
$\dn$ mesons are produced in the entire $x_p$ range from $0$ to
$1$. The visible $\ee\to\bbbar\to\dn\,X$ cross section for the i-th
$\ebb$ bin is calculated as 
\begin{equation}
  \sigv^{\mathrm{(i)}}(\ee\to\bbbar\to\dn\,X)\propto\frac{\NiL-\rco\,\NiH}{\NiH\,\ecm^2}.
\end{equation}
Here the $\dn$ yield in the $x_p>0.55$ region is used both to subtract
the $\ee\to\ccbar\to\dn\,X$ continuum (the factor $\rco$ is the ratio
of the continuum yields in the $x_p<0.5$ and $x_p>0.55$ regions) and
for normalization (the factor $\ecm^2$ accounts for the fact that the
cross section of the continuum process is proportional to $1/\ecm^2$).

\begin{figure}[htbp]
  \centering
  \includegraphics[width=0.49\linewidth]{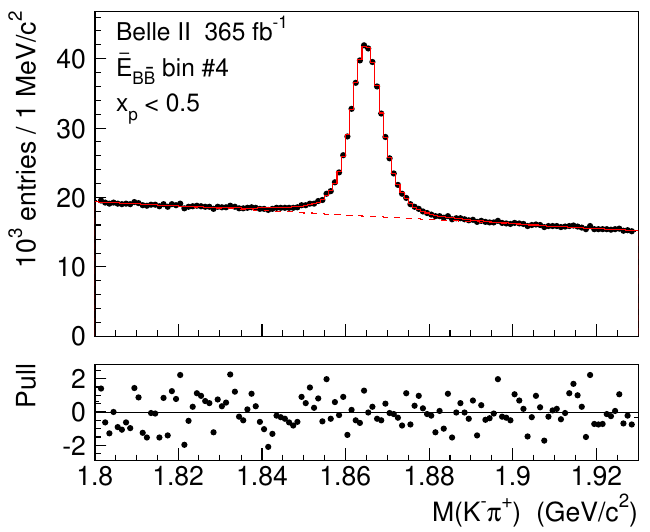}\hfill
  \includegraphics[width=0.49\linewidth]{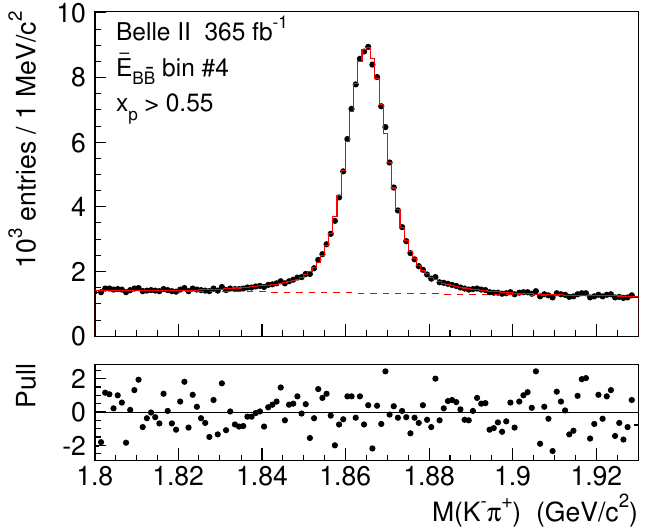}
  \caption{The $M(K^-\pi^+)$ distributions for $D^0$ candidates with
    $x_p<0.5$ (left) and $x_p>0.55$ (right) in the 4th $\ebb$ bin. The
    solid histogram is the result of the fit; the dashed histogram is
    the background component of the fit.}
  \label{m_d0_total_xp_lo_050523}
\end{figure}

We fit the $M(K^-\pi^+)$ distributions for the $x_p<0.5$ and
$x_p>0.55$ regions for the entire $\ebb$ interval with a sum of three
Gaussian functions and a first order polynomial.
We then fit the $M(K^-\pi^+)$ distributions in $\ebb$ bins by fixing
the parameters of the Gaussian functions and introducing a common
shift and broadening factor that are free in the fit. The parameters
of the background function are also allowed to float.
An example of the fit to the $M(K^-\pi^+)$ distributions in the 4th
$\ebb$ bin is shown in figure~\ref{m_d0_total_xp_lo_050523} (the $\ebb$
bins are shown in figure~\ref{ecm_belle_270225}).

\begin{figure}[htbp]
\centering
\includegraphics[width=0.5\linewidth]{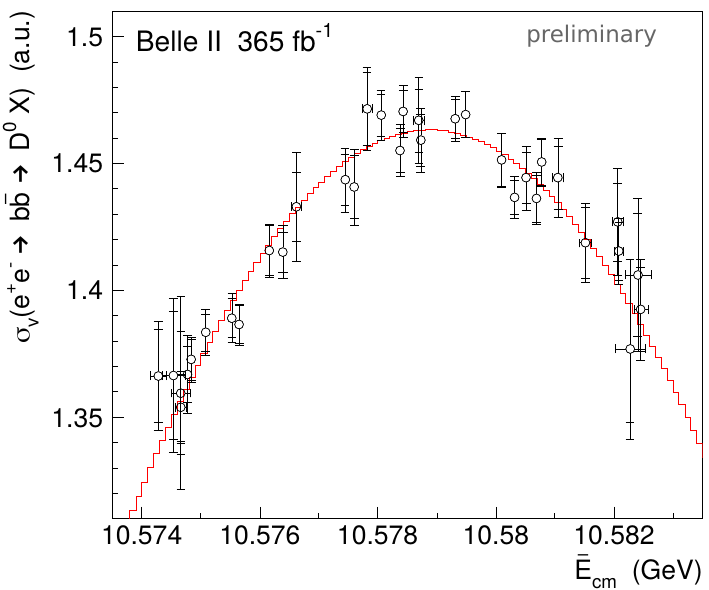}
\caption{ Visible $\ee\to\bbbar\to{}\dnx$ cross section versus
  $\aecm$. The open circles with error bars show the direct
  measurement performed using Belle~II data; the inner and outer error
  bars show statistical and total uncorrelated uncertainties,
  respectively. The curve is the result of the combined fit described
  in section~\ref{sec:results}. The vertical scale is in arbitrary
  units. }
\label{xs_ds_280324}
\end{figure}

The factor $\rco$ is determined using fits to the $M(K^-\pi^+)$
distributions in an off-resonance data sample that has an integrated
luminosity of $43\,\fb$, collected at $\ecm=10.52\,\gev$. We find
$\rco=0.608\pm0.006$.
The result for the visible $\ee\to\bbbar\to{}\dnx$ cross section in
arbitrary units is shown in figure~\ref{xs_ds_280324}.

In the combined fit, the visible $\ee\to\bbbar\to{}\dnx$ cross section
is described by the function
\begin{align*}
\sigv(\ee\to\bbbar\to{}\dnx)= & \;\sigv(\ee\to\bpbm)\;2\,\br(\bp\to{}\dnx)\; + \\
        & \;\sigv(\ee\to\bnbn)\;\;2\,\br(\bn\to{}\dnx).
\end{align*}
We use the world-average values~\cite{ParticleDataGroup:2024cfk}
\begin{align}
  \br(\bp\to{}\dnx)=(87.6\pm4.1)\%, \label{eq:bf_bp_d0} \\
  \br(\bn\to{}\dnx)=(55.5\pm3.2)\%. \label{eq:bf_bn_d0}  
\end{align}
Since the vertical scale in figure~\ref{xs_ds_280324} is arbitrary, we
multiply the fit function by a normalization parameter, which is free
in the fit.

To account for the uncertainties in $\aecm$, $\delta(\ecm)$, we use
the formula 
\begin{equation}
  \delta{}\sigv = |f^{\prime}(\ecm)| \, \delta(\ecm),
  \label{eq:sigma_xy}
\end{equation}
where $f^{\prime}(\ecm)$ is the derivative of the
$\sigv(\ee\to\bbbar\to{}\dnx)$ dependence on $\ecm$.
At the boundaries of the $\aecm$ range, where the values of
$|f^{\prime}(\ecm)|$ are high, the contribution of the $\aecm$
uncertainty is up to 40\% of the statistical uncertainty in the
cross section.
%
In the case of the $\rnpv$ energy dependence described in
section~\ref{sec:rat_vs_ecm}, the contribution of the uncertainty in
$\aecm$ is negligibly small.

To estimate the systematic uncertainty associated with the signal
shape, we introduce additional corrections: a shift and broadening for
the sum of the second and third Gaussians in the parameterization of
the signal shape. These two Gaussians account for about 50\% of the
signal yield. To estimate the systematic uncertainty associated with
the background modeling, we change the order of the polynomial that
parameterizes the background shape from first to second. We repeat the
fits and consider deviations as symmetric systematic uncertainties,
which are assumed to be uncorrelated between $\ebb$ bins.  The
uncertainties from the two sources are comparable.
To obtain the total uncertainty, we add in quadrature the
statistical uncertainty, the contribution due to the uncertainty in
$\ecm$, and the systematic contributions. The total and statistical
uncertainties are shown in figure~\ref{xs_ds_280324}.

The $\ee\to\bbbar\to{}\dnx$ cross-section measurements at Belle
cover a smaller energy interval compared to Belle~II. They agree with
the expectations based on the combined fit but do not provide a
significant constraint on the cross-section peak position.
For simplicity, we do not include the $\ee\to\bbbar\to{}\dnx$
cross-section measurement at Belle in the combined fit.

\section{Combined fit and its results}
\label{sec:results}

We perform a simultaneous fit to the distributions shown in
figures~\ref{mbc_y4s_bp_belle_280324}, \ref{mbc_y4s_bp_belle2_280324},
\ref{rat_vs_ecm_small_280324}, \ref{xsec_y4s_280324} and
\ref{xs_ds_280324}. The corresponding fit functions are described in
the previous sections.
The fits to $\mt$ distributions are binned likelihood ($-2\ln{}L$)
fits, while the fits to the energy dependence of the $R_b$, $B^0/B^+$
cross-section ratio and $\sigma(\ee\to\bbbar\to{}\dnx)$ are $\chi^2$
fits. The $-2\ln{}L$ and $\chi^2$ values are added together and the
sum is passed to the fit for minimization.
The results are shown in these figures and in 
tables~\ref{tab:y4s_fit_results_m_e}, 
\ref{tab:y4s_fit_results_yields} and \ref{tab:y4s_fit_results_pbg}.
Throughout this paper, the first uncertainty is statistical and the
second, if present, is systematic.
\begin{table}[htbp]
  \caption{ Fit results for $\dm$ and the c.m.\ energy-related
    variables. }
  \renewcommand*{\arraystretch}{1.1}
  \label{tab:y4s_fit_results_m_e}
  \centering
  \begin{tabular}{@{}lc@{}} \toprule
    $\dm$             &  $(0.495\pm0.024\pm0.005)\,\mevm$ \\
    \midrule
    Belle & \\
    \;\;\;\;\;$\ez$       &  $(10579.05\pm0.04\pm0.18)\,\mev$ \\
    \;\;\;\;\;$\spread$   &  $(5.35\pm0.07\pm0.09)\,\mev$ \\
    Belle~II & \\
    \;\;\;\;\;$\ez$       &  $(10578.19\pm0.05\pm0.22)\,\mev$ \\
    \;\;\;\;\;$\spread$   &  $(5.24\pm0.08\pm0.15)\,\mev$ \\
    $\deb$            &  $(-1.27\pm0.14\pm0.22)\,\mev$ \\
    \bottomrule
  \end{tabular}
\end{table}
\begin{table}[htbp]
  \caption{ Fit results for the $\bp$ and $\bn$ signal yields and the efficiency ratio. }
  \renewcommand*{\arraystretch}{1.2}
  \label{tab:y4s_fit_results_yields}
  \centering
  \begin{tabular}{@{}lcc@{}} \toprule
    & Belle & Belle~II \\
    \midrule
    $N_{\bp}+N_{\bn}$    & $(587.6\pm0.8)\cdot10^3$ & $(458.2\pm0.7)\cdot10^3$ \\
    $N_{\bn}\;/\;N_{\bp}$ & $0.789\pm0.002$ & $0.895\pm0.003$ \\
    $r_{\varepsilon}$     & $0.830\pm0.002$ & $0.949\pm0.003$ \\
    \bottomrule
  \end{tabular}
\end{table}
\begin{table}[htbp]
  \caption{ Fit results for the peaking background corrections. }
  \renewcommand*{\arraystretch}{1.2}
  \label{tab:y4s_fit_results_pbg}
  \centering
  \begin{tabular}{@{}lcccc@{}} \toprule
    & \multicolumn{2}{c}{Belle} & \multicolumn{2}{c}{Belle~II} \\
    & $\bp$ & $\bn$ & $\bp$ & $\bn$ \\
    \midrule
    $\nr_3$          & $1.22\pm0.07$ & $1.20\pm0.13$ & $1.03\pm0.06$ & $1.13\pm0.09$ \\
    $\sh_3$, $\mevc$ & $-5.8\pm2.7$  & $-3.0\pm4.4$  & $15.0\pm3.1$  & $17.0\pm3.9$ \\
    $\ff_3$          & $0.84\pm0.06$ & $1.02\pm0.12$ & $0.64\pm0.04$ & $0.53\pm0.05$ \\
    \bottomrule
  \end{tabular}
\end{table}
The tables show the results for 24 fit parameters. In addition, there
are 12 free parameters describing the $R_b$ energy dependence, 3
parameters describing $\rnp(\ecm)$, 20 parameters describing the
yields and shapes of smooth background in the $\mt$ distributions, and
one more parameter describing the normalization of the $\sigma(\dnx)$
measurement. Thus, there are a total of 60 free parameters in the fit.
All distributions are described well by the fit functions. The results
for $\ez$ and $\spread$ at Belle, and $\deb$ are in agreement with the
Belle-only results~\cite{Belle:2021lzm}.
The fitted values of $\sigma(b\bar{b})$ and $\rnp$ as functions of
$\ecm$ with a step size of $0.1\,\mev$ are given in
appendix~\ref{app:xsecb_rat}. We also provide fitted values for
pseudo-experiments and fit modifications for the main sources of
systematic uncertainty. This information can be used to perform energy
calibration near the $\Ufo$ peak or for phenomenological studies.

\section{Systematic uncertainties and additional studies}
\label{sec:syst}

The contributions to the systematic uncertainty in the fit results are
shown in table~\ref{tab:dm_syst}.
\begin{table}[htbp]
\caption{ Systematic uncertainties from various sources in $\dm$ (in
  $\mevm$), $\ez$ and $\spread$ (for Belle and Belle~II), and $\deb$
  (in $\mev$). }
\renewcommand*{\arraystretch}{1.1}
\label{tab:dm_syst}
\centering
\begin{tabular}{@{}lcccccc@{}} \toprule
  & \multirow{2}{*}{$\dm$} & \multicolumn{2}{c}{Belle} & \multicolumn{2}{c}{Belle II} & \multirow{2}{*}{$\deb$} \\
  &  & $\ez$ & $\spread$ & $\ez$ & $\spread$ &  \\
  \midrule
  $\bp$ mass                          & 0.001 & 0.160 & 0.011 & 0.160 & 0.013 & 0.163 \\
  Absolute value of $\rnpv$           & 0.002 & 0.007 & 0.007 & 0.008 & 0.008 & 0.009 \\
  $\br(B\to \dnx)$                    & 0.001 & 0.005 & 0.001 & 0.005 & 0.002 & 0.008 \\
  $R_b$ vs.\ $\ecm$ parameterization  & 0.000 & 0.040 & 0.040 & 0.040 & 0.052 & 0.071 \\
  $\rnp$ vs.\ $\ecm$ parameterization & 0.001 & 0.048 & 0.046 & 0.044 & 0.060 & 0.086 \\
  Momentum scale                      & 0.001 & 0.002 & 0.000 & 0.001 & 0.000 & 0.002 \\
  Momentum resolution                 & 0.002 & 0.007 & 0.005 & 0.043 & 0.045 & 0.012 \\
  Peaking background                  & 0.003 & 0.023 & 0.053 & 0.127 & 0.114 & 0.051 \\
  Smooth background                   & 0.000 & 0.047 & 0.029 & 0.042 & 0.037 & 0.078 \\
  Binning                             & 0.002 & 0.004 & 0.004 & 0.001 & 0.006 & 0.003 \\
  \midrule
  Total                               & 0.005 & 0.180 & 0.087 & 0.221 & 0.151 & 0.219 \\
\bottomrule
\end{tabular}
\end{table}
We consider the following sources of systematics uncertainties.
\begin{enumerate}
\item {\bf\boldmath Uncertainty in the $\bp$ mass.} For the $\bp$ mass, we
  use the PDG average (not the PDG fit):
  $m(\bp)=(5279.42\pm0.08)\,\mevm$~\cite{ParticleDataGroup:2024cfk}. We
  vary its value by $\pm1$ standard deviation and repeat the analysis.
  Here and in all cases described below, we estimate the systematic
  uncertainty as the maximum deviation of the fit result, assuming
  that the systematic uncertainties are symmetric.
\item {\bf\boldmath Uncertainty in the absolute value of $\rnpv$.} We
  vary the world-average value
  $\rnpv=0.951\pm0.028$~\cite{HeavyFlavorAveragingGroupHFLAV:2024ctg}
  measured at the $\Ufo$ peak by its uncertainty.
\item {\bf\boldmath Uncertainty in $\br(B\to \dnx)$.} We vary the
  values of $\br(\bp\to \dnx)=(87.6\pm4.1)\%$ and
  $\br(\bn\to{}\dnx)=(55.5\pm3.2)\%$~\cite{ParticleDataGroup:2024cfk}
  by their uncertainties. The contributions from $\bn$ and $\bp$ are added
  in quadrature.
\item {\bf\boldmath Parameterization of $R_b$ energy dependence.} In
  the default fit, the energy dependence of $R_b$ is parameterized by
  an 11th order Chebyshev polynomial. We also fit with 12th and 13th
  orders as a cross-check.
\item {\bf\boldmath Parameterization of $\rnp$ energy dependence.} We
  find that a second order Chebyshev polynomial is sufficient to
  describe the $\rnp$ energy dependence; increasing the order gives a
  very small reduction of the overall $\chi^2-2\ln{}L$ of the fit. We
  consider polynomial orders up to 10. The results of the fits with
  different polynomial orders for the ratio of visible and dressed
  cross sections are shown in
  figure~\ref{rat_vs_ecm_poly_orders_140324}.
\begin{figure}[thb]
\centering
\includegraphics[width=0.49\linewidth]{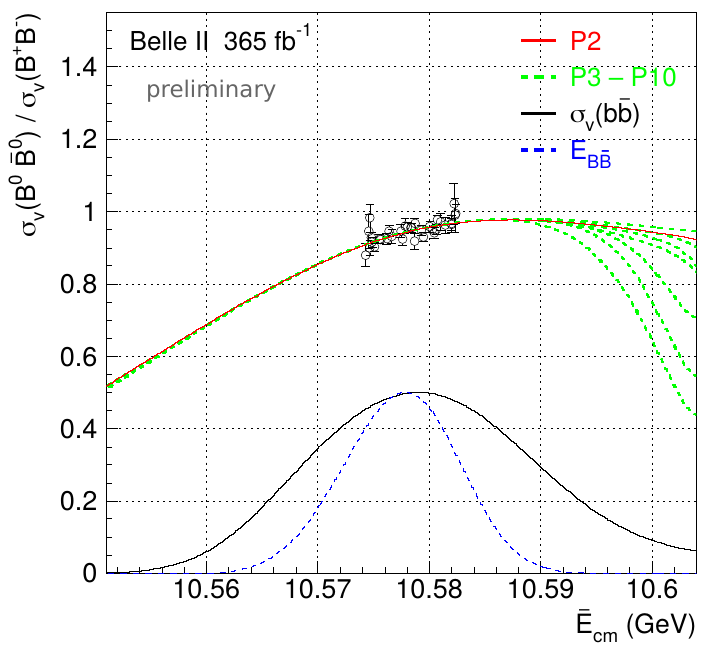}\hfill
\includegraphics[width=0.49\linewidth]{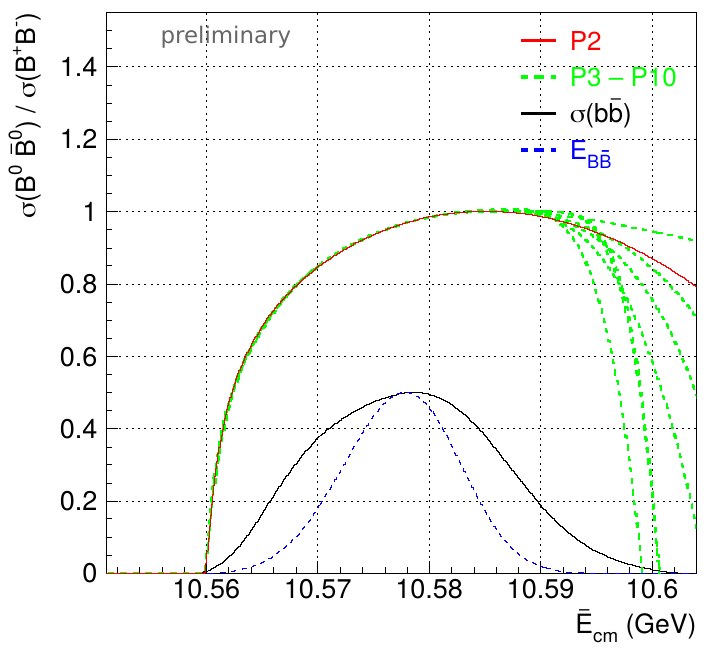}
\caption{ (Left) the ratio of visible cross sections
  $\sigv(\ee\to\bnbn)\,/\,\sigv(\ee\to\bpbm)$ versus energy. The open
  circles with error bars show the direct measurement using Belle~II
  data; the red solid curve is the result of the nominal fit; the
  green dashed curves correspond to the orders of the polynomial
  describing the $\rnp$ energy dependence from three to 10; the black
  solid curve shows the energy dependence of the visible $\ee\to\bb$
  cross section; the blue dashed curve shows the distribution of the
  number of produced $\bb$ pairs. Normalization (vertical scale) of
  the latter two curves is arbitrary. (Right) the same curves as on
  the left panel, except that they correspond to dressed cross
  sections before accounting for the energy spread. }
\label{rat_vs_ecm_poly_orders_140324}
\end{figure}
In the same figure, we also show the energy dependence of the
$\ee\to\bbbar$ cross section and the distribution of the number of
produced $\bb$ pairs, which corresponds to the first line of
eq.~\eqref{eq:scheme}.
The curves corresponding to different polynomial orders coincide in
the energy region covered by $E_{B\bar{B}}$, but begin to diverge
outside this region. This is because the $\mt$ distributions constrain
the cross-section shapes in the region covered by $E_{B\bar{B}}$.
\item {\bf Uncertainty in the momentum scale.} We find that the $\DE'$
  peak positions differ slightly between simulation and data. We
  introduce momentum-scale corrections in the simulation to reproduce
  the peak positions in the data.
\item {\bf Uncertainty in the momentum resolution.} We vary the correction
  factors $\ff_1$ for the width of the momentum resolution function by
  $\pm3\%$, which is the relative difference of the width correction
  factors for the $\DE'$ and momentum resolutions
  (section~\ref{sec:mbc_fit}).
\item {\bf Modelling of peaking background.} We have two types of
  peaking background: in the $\DE'$ signal region and in the $\DE'$
  sideband. The shapes of these backgrounds and their yield relative
  to the signal events are determined from simulation. In the default
  model, we introduce correction parameters for the peaking component
  in the sideband, $\nr_3$, $\sh_3$ and $\ff_3$ (for yield, peak
  position and broadening), while corrections for the peaking
  background in the $\DE'$ signal region are fixed at the values
  $\nr_2=1$, $\sh_2=0$ and $\ff_2=1$. We consider two alternative
  peaking background models. In the first, we assume $\nr_2=\nr_3$,
  $\sh_2=\sh_3$ and $\ff_2=\ff_3$, and treat these as free parameters
  in the fit. In the second model, the parameters $\nr_2$, $\sh_2$ and
  $\ff_2$ are varied in the fit independently.
  The peaking background is enhanced in channels with $D_{(s)}^*$ due
  to soft pions or $\gamma$ from $D_{(s)}^*$ decays. We exclude all
  such channels when plotting the $\mt$ distributions. We lose $1/2$
  ($1/3$) of events for $\bn$ ($\bp$), while the peaking-background
  fraction is reduced by about a factor of 2.5. We then repeat the
  combined fit. In all cases, the changes in $\dm$ are small
  (table~\ref{tab:dm_syst}). 
\item {\bf Modelling of smooth background.} The smooth background is
  described by a second order polynomial multiplied by a threshold
  function; we increase the order of the polynomial by one.
\item {\bf Binning.} We shift the binning in $\mt$ distributions
  (figures~\ref{mbc_y4s_bp_belle_280324} and
  \ref{mbc_y4s_bp_belle2_280324}) and in $\ebb$ distributions
  (figure~\ref{ecm_belle_270225}) by half the bin size.
\end{enumerate}

To test how the various distributions used in the combined fit affect
the precision in $\dm$, we perform the fit excluding some
distributions. The uncertainty in $\dm$ changes very little if we
exclude the $\mt$ distributions of Belle or Belle~II. 
In contrast, the precision in $\dm$ deteriorates significantly if
either Belle or Belle~II data on $\rnpv(\ecm)$ are excluded. If the
$\rnpv$ data of both experiments are excluded, the uncertainty in
$\dm$ increases to $0.10\,\mevm$. Thus, it is the slope of $\rnp$
versus $\ecm$ that determines the precision in $\dm$. 
If we exclude the $\sigma(\ee\to\dnx)$ measurement, the statistical
uncertainty of $\ez$ in Belle and Belle~II increases by a factor of 4;
the statistical uncertainty of $\deb$ also increases, while all other
results remain unchanged. In all cases, we find that the results with
some measurements excluded are consistent with the nominal fit. 

The measured value of $\dm$, $(0.495\pm0.024\pm0.005)\,\mevm$, differs
significantly from the BaBar result,
$(0.33\pm0.05\pm0.03)\,\mevm$~\cite{BaBar:2008ikz}. To study the
origin of this difference, we perform the combined fit using, as in
ref.~\cite{BaBar:2008ikz}, the phase-space hypothesis,
$\rnp=(p_{\bn}/p_{\bp})^3$. The change in $\chi^2-2\ln{}L$ of the fit
is $+113.1$ with the dominant contribution, $+95.5$, coming from the
$\mt$ distributions (the signal fit function is too broad for $\bn$
and too narrow for $\bp$). Based on Wilks' theorem, the phase-space
hypothesis is disfavored at about the $10\,\sigma$ level.\footnote{The
  difference in the number of degrees of freedom of the two hypotheses
  is three; it is the number of free parameters of $P_2(\ecm)$ in
  eq.~\eqref{eq:ee_to_bnbn}.} The value of $\dm$ from this phase-space
hypothesis fit is $(0.386\pm0.006)\,\mevm$, which is consistent with
the BaBar result~\cite{BaBar:2008ikz}. Thus, the difference in $\dm$
(table~\ref{tab:y4s_fit_results_m_e}) with respect to BaBar is
probably due to the fact that BaBar used the phase-space hypothesis
for the energy dependence of $\rnp$, which we find to be excluded. 

Comparing the left and right panels in
figure~\ref{rat_vs_ecm_poly_orders_140324}, we see that the ratio of
dressed cross sections $\rnp$ reaches 1.0 while the ratio of visible
cross sections $\rnpv$ remains a few per cent below one. This shows
the importance of the ISR and energy-spread effects for the absolute
value of the cross-section ratio.

\section{\boldmath Measurement of the ratio of dressed $\bnbn$ to $\bpbm$ cross sections in a wide energy range}
\label{sec:r_vs_ecm_wide}

\begin{figure}[htbp]
\centering
\includegraphics[width=0.48\linewidth]{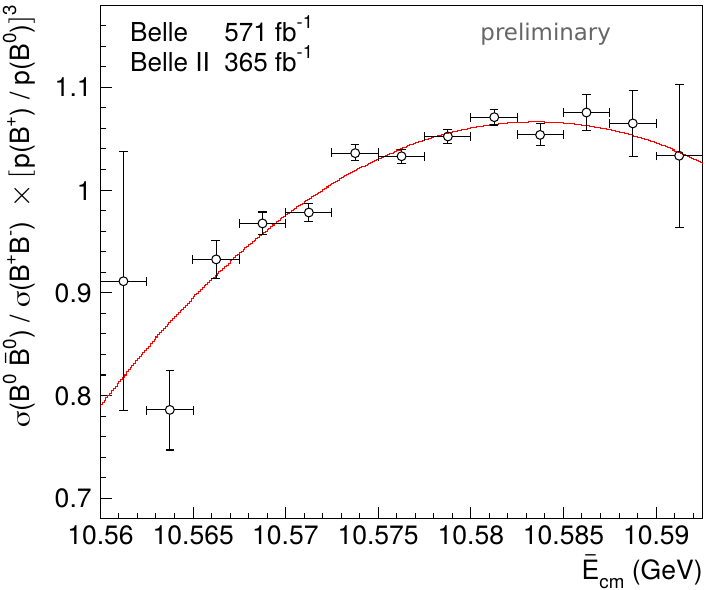}\hfill
\includegraphics[width=0.48\linewidth]{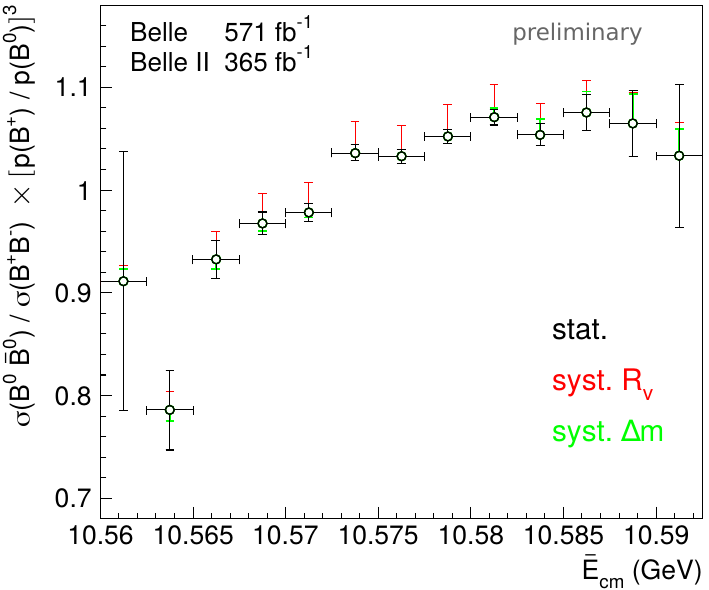}
\caption{ Energy dependence of the ratio of dressed $\ee\to\bnbn$ and
  $\ee\to\bpbm$ cross sections divided by the phase-space factor
  $(p_{\bn}/p_{\bp})^3$. The open circles with error bars show the
  results of the modified combined fit; the red curve (left) shows the
  original second-order polynomial; the red and green error bars
  (right) show the systematic shifts due to variation of,
  respectively, $\rnpv$ and $\dm$ by $+1$ standard deviation. The
  shifts due to $-1\,\sigma$ variations are not shown to make the
  signs of the shifts clear. }
\label{r_over_phsp_points_p2_280425}
\end{figure}

The combined fit constrains the energy dependence of the dressed
cross-section ratio $\rnp$ in a much wider range than that covered by
direct measurement (figure~\ref{rat_vs_ecm_poly_orders_140324}).
This is because the $\mt$ spectra of $\bn$ and $\bp$ candidates
contain information about the shapes of $\bnbn$ and $\bpbm$ cross
sections, as discussed in section~\ref{sec:intro}. 
To provide information about the energy dependence of $\rnp$ for
phenomenological studies, we measure $\rnp$ in $\ecm$ bins using the
combined fit. The idea is similar to the Argand plot measurement in
amplitude analyses~\cite{LHCb:2014zfx,Belle:2014nuw}. 
The dressed cross-section ratio corrected for the phase-space factor,
$\rnp(\ecm)/(p_{\bn}/p_{\bp})^3$, is described by a second order
polynomial $P_2(\ecm)$ in the combined fit
(eq.~\eqref{eq:ee_to_bnbn}). 
We modify the combined fit, and instead of $P_2(\ecm)$, we use a
piecewise step function where the height of each step is determined
from the fit. 
The quantity $\rnp(\ecm)/(p_{\bn}/p_{\bp})^3$ is particularly
convenient for this approach, since it changes relatively slowly with
energy. 
We chose bins of width $2.5\,\mev$, which results in 13 bins in the
energy range $10.56-10.5925\,\gev$. For the small fraction of events
outside this range, we use the values in the first or last bin. 
The original polynomial $P_2(\ecm)$ and the results of the modified
fit are shown in figure~\ref{r_over_phsp_points_p2_280425}. 

To better control the correlations among the
$\rnp(\ecm)/(p_{\bn}/p_{\bp})^3$ measurements, the $\dm$ value is
fixed in the combined fit, and the uncertainty in $\dm$ is taken into
account as systematic uncertainty. 
We also consider the absolute value of $\rnpv$
(eq.~\eqref{eq:abs_fnp}) as a source of correlated systematic
uncertainty. Other sources of systematic uncertainty
(table~\ref{tab:dm_syst}) have negligible influence on $\rnp$. The
values of $\rnp/(p_{\bn}/p_{\bp})^3$ with statistical and systematic
uncertainties for different $\ecm$ bins are given in
table~\ref{tab:rat}. 
\begin{table}[htbp]
\caption{Measured values of the dressed cross-section ratio corrected
  for the phase-space factor, $\rnp/(p_{\bn}/p_{\bp})^3$, at various
  energies. Shown are the $\ecm$ interval (in $\mev$), the central
  value of $\rnp/(p_{\bn}/p_{\bp})^3$ with its statistical
  uncertainty, and the systematic shifts from changing the $\rnpv$ and
  $\dm$ values by $\pm1\,\sigma$.} \renewcommand*{\arraystretch}{1.1}
\label{tab:rat}
\centering
\begin{tabular}{@{}rrrrr@{}} \toprule
  N & $\ecm$ interval & $\rnp/(p_{\bn}/p_{\bp})^3$ & syst. $\rnpv$ & syst. $\dm$ \\
  \midrule
  1 & $10561.25\pm1.25$ & $0.9111\pm0.1261$ & $\pm0.0152$ & $\pm0.0121$ \\
  2 & $10563.75\pm1.25$ & $0.7858\pm0.0389$ & $\pm0.0185$ & $\mp0.0104$ \\
  3 & $10566.25\pm1.25$ & $0.9323\pm0.0183$ & $\pm0.0274$ & $\mp0.0088$ \\
  4 & $10568.75\pm1.25$ & $0.9674\pm0.0112$ & $\pm0.0292$ & $\mp0.0068$ \\
  5 & $10571.25\pm1.25$ & $0.9779\pm0.0090$ & $\pm0.0291$ & $\mp0.0049$ \\
  6 & $10573.75\pm1.25$ & $1.0362\pm0.0077$ & $\pm0.0303$ & $\mp0.0029$ \\
  7 & $10576.25\pm1.25$ & $1.0327\pm0.0067$ & $\pm0.0302$ & $\pm0.0014$ \\
  8 & $10578.75\pm1.25$ & $1.0524\pm0.0066$ & $\pm0.0311$ & $\pm0.0041$ \\
  9 & $10581.25\pm1.25$ & $1.0709\pm0.0076$ & $\pm0.0318$ & $\pm0.0094$ \\
 10 & $10583.75\pm1.25$ & $1.0537\pm0.0108$ & $\pm0.0311$ & $\pm0.0152$ \\
 11 & $10586.25\pm1.25$ & $1.0755\pm0.0176$ & $\pm0.0312$ & $\pm0.0202$ \\
 12 & $10588.75\pm1.25$ & $1.0647\pm0.0320$ & $\pm0.0305$ & $\pm0.0292$ \\
 13 & $10591.25\pm1.25$ & $1.0333\pm0.0692$ & $\pm0.0320$ & $\pm0.0260$ \\
\bottomrule
\end{tabular}
\end{table}
The correlation matrix for the measured $\rnp/(p_{\bn}/p_{\bp})^3$
values is shown in table~\ref{tab:cor_matrix}. 
\begin{table}[htbp]
  \caption{ Correlation matrix for the 13 measured values of
    $\rnp/(p_{\bn}/p_{\bp})^3$ shown in
    figure~\ref{r_over_phsp_points_p2_280425}. } 
\renewcommand*{\arraystretch}{1.1}
\label{tab:cor_matrix}
\centering
\scriptsize
\begin{tabular}{@{}rrrrrrrrrrrrr@{}} \toprule
 $1.00$ & $-0.24$ & $0.16$ & $-0.01$ & $0.03$ & $-0.10$ & $0.07$ & $-0.05$ & $0.00$ & $-0.02$ & $0.01$ & $0.03$ & $0.07$ \\ 
        & $1.00$ & $-0.21$ & $0.11$ & $-0.07$ & $-0.04$ & $-0.02$ & $-0.02$ & $-0.05$ & $0.02$ & $-0.02$ & $-0.01$ & $0.01$ \\ 
        &         & $1.00$ & $-0.22$ & $0.08$ & $-0.12$ & $-0.03$ & $-0.01$ & $-0.03$ & $-0.01$ & $-0.02$ & $-0.04$ & $0.11$ \\ 
        &         &         & $1.00$ & $-0.23$ & $0.10$ & $-0.14$ & $-0.02$ & $-0.06$ & $0.02$ & $-0.03$ & $-0.02$ & $-0.01$ \\ 
        &         &         &         & $1.00$ & $-0.43$ & $0.18$ & $-0.17$ & $-0.00$ & $-0.07$ & $-0.01$ & $-0.01$ & $-0.00$ \\ 
        &         &         &         &         & $1.00$ & $-0.45$ & $0.16$ & $-0.18$ & $-0.00$ & $-0.03$ & $0.00$ & $-0.07$ \\ 
        &         &         &         &         &         & $1.00$ & $-0.52$ & $0.15$ & $-0.18$ & $-0.03$ & $-0.01$ & $-0.05$ \\ 
        &         &         &         &         &         &         & $1.00$ & $-0.54$ & $0.19$ & $-0.16$ & $0.01$ & $0.01$ \\ 
        &         &         &         &         &         &         &         & $1.00$ & $-0.50$ & $0.21$ & $-0.13$ & $-0.00$ \\ 
        &         &         &         &         &         &         &         &         & $1.00$ & $-0.54$ & $0.24$ & $-0.20$ \\ 
        &         &         &         &         &         &         &         &         &         & $1.00$ & $-0.51$ & $0.21$ \\ 
        &         &         &         &         &         &         &         &         &         &         & $1.00$ & $-0.39$ \\ 
        &         &         &         &         &         &         &         &         &         &         &         & $1.00$ \\ 
\bottomrule
\end{tabular}
\end{table}

\section{\boldmath Phenomenological analysis of $\rnp$ versus $\ecm$}
\label{sec:pheno}

To fit the data shown in figure~\ref{r_over_phsp_points_p2_280425}, we
use the phenomenological model of
refs.~\cite{Dubynskiy:2007xw,Voloshin:2018ejo}. 
In this model, the strong interaction between $\bb$ mesons is
described by a potential that has a constant value ($V_1$ for the
isovector component) inside an interaction region of radius $a$ and
zero outside. At distances $r<a$ the wave functions of $B$ and
$\bar{B}$ overlap, so that it is impossible to distinguish the $\bnbn$
and $\bpbm$ pairs. Therefore, at $r<a$ the electromagnetic interaction
is switched off in the model. 
The model has two parameters: the interaction range $a$ and the $\bb$
scattering phase in the isovector channel $\delta_1$. 
The scattering phase is proportional to the isovector potential $V_1$
and varies with $\ecm$; as a fit parameter, we take the $\delta_1$
value at $\ecm=10.58\,\gev$. 
The model is valid when $|\rnp-1|\ll{}1$, so in the fitting we exclude
the two lowest energy points. We also exclude the highest energy point
because it contains all events above $10.59\,\gev$, and hence its
argument is not well defined; moreover, the corresponding uncertainty
in $\rnp/(p_{\bn}/p_{\bp})^3$ is poor. 
Detailed information about the model, as well as the expression for
$\rnp$, are given in appendix~\ref{app:pheno}. 

We define the pull of each experimental measurement, $x_i\pm\sigma_i$
($i=1,10$), relative to the fitted value, $\mu_i$, as 
\begin{equation}
  p_i = (x_i + \sum_{j=1}^2\xi_j\,\Delta{}x_{ji} - \mu_i)/\sigma_i\,.
  \label{eq:pool}
\end{equation}
Here the second term takes into account systematic uncertainties due
to external parameters: $\Delta{}x_{ji}$ is the signed deviation of
$x_i$ when the $j$-th external parameter changes by $+1$ standard
deviation ($j$ runs over the $\rnpv$ and $\dm$ contributions); $\xi_j$
is a nuisance parameter that is free in the fit. 
The $\chi^2$ of the fit is defined 
as~\cite{HeavyFlavorAveragingGroupHFLAV:2024ctg} 
\begin{equation}
  \chi^2 = \sum_{i=1}^{10}C^{-1}_{ij}p_ip_j + \sum_{j=1}^2\xi_j^2\,,
  \label{eq:chisq}
\end{equation}
where $C^{-1}$ is the inverse of the correlation matrix of
measurements $x_i$ used in the fit. 

\begin{figure}[htbp]
\centering
\includegraphics[width=0.49\linewidth]{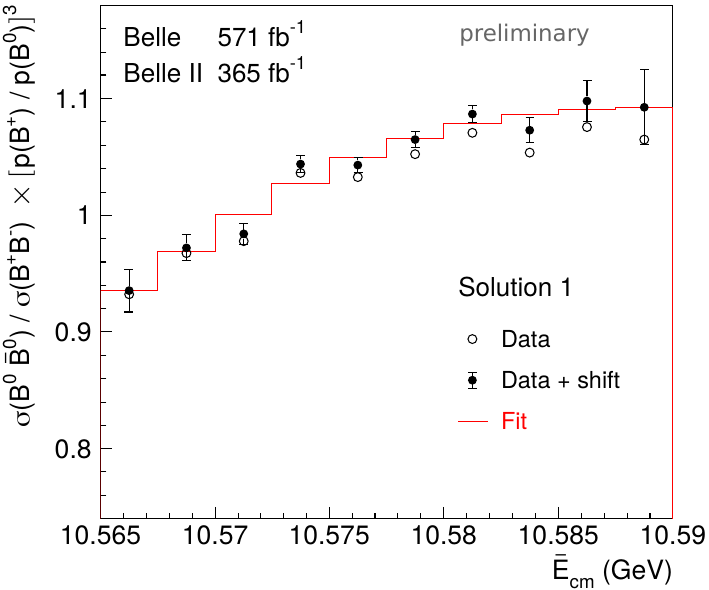}\hfill
\includegraphics[width=0.49\linewidth]{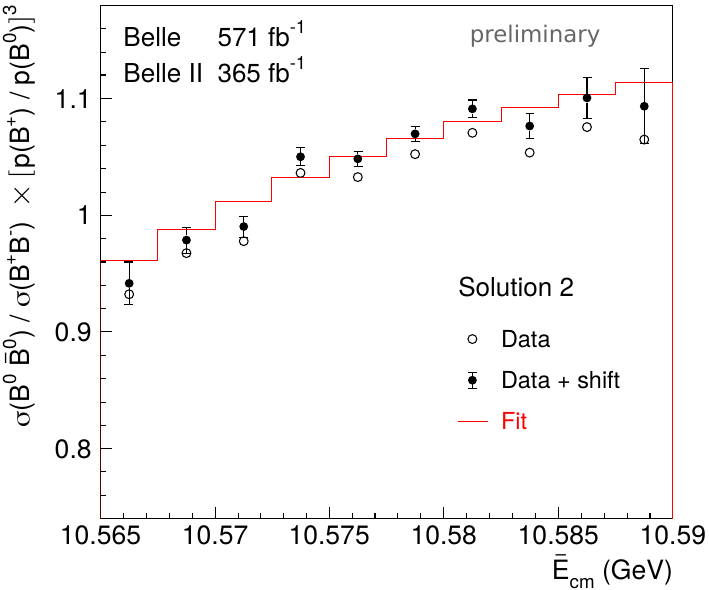}
\caption{ Energy dependence of the ratio of dressed $\ee\to\bnbn$ and
  $\ee\to\bpbm$ cross sections divided by the phase-space factor
  $(p_{\bn}/p_{\bp})^3$. The black dots with error bars are the
  measured values shifted by the correlated systematic uncertainty
  (the second term in eq.~\eqref{eq:pool}) according to the fit
  results; the open black circles are the measured values before the
  shift. The red histogram shows the result of the fit to the
  phenomenological model. The left and right panels show solutions 1
  and 2, respectively. }
\label{r_over_phsp_solution1_280425}
\end{figure}

\begin{figure}[htbp]
\centering
\includegraphics[width=0.49\linewidth]{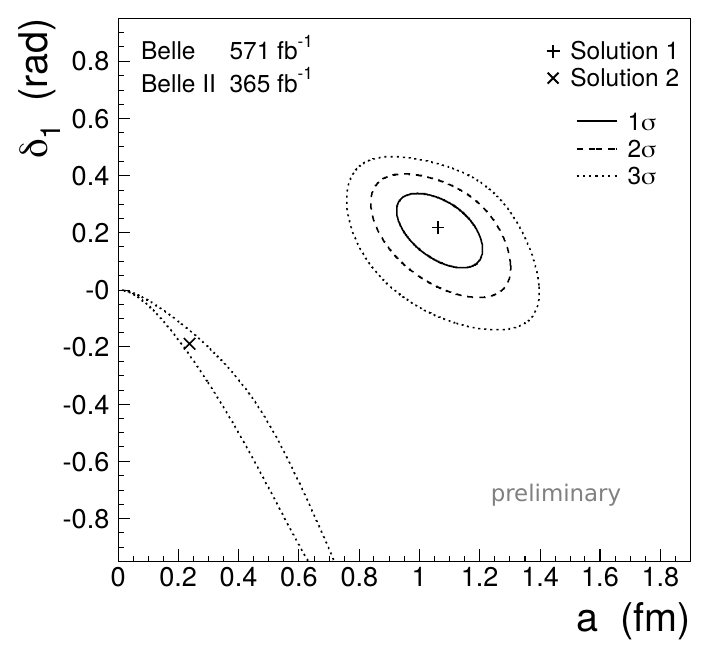}
\caption{ The $(a, \delta_1)$ plane. Here $a$ and $\delta_1$ are
  phenomenological parameters of an isovector $B\bar{B}$ potential
  described in the text.}
\label{aa_vs_ph_150325_rel_dm}
\end{figure}

We perform a 2D scan in the plane of variables $a$ and $\delta_1$. At
each scan point ($a$ and $\delta_1$ are fixed), we perform a fit with
two free parameters, $\xi_{\rnpv}$ and $\xi_{\dm}$. The fit results
for the best solution are shown in
figure~\ref{r_over_phsp_solution1_280425} and
table~\ref{tab:solution}. 
\begin{table}[htbp]
\caption{ Parameters of two solutions in the $(a,\delta_1)$ plane.}
\renewcommand*{\arraystretch}{1.3}
\label{tab:solution}
\centering
\begin{tabular}{@{}lcc@{}} \toprule
  & Solution 1 & Solution 2 \\
  \midrule
  $a$, fm         & $1.062\pm0.094$ & $0.24^{+0.24}_{-0.15}$ \\
  $\delta_1$, rad & $0.218^{+0.076}_{-0.097}$ & $-0.19^{+0.16}_{-0.33}$ \\
  $\xi_{\rnpv}$   & $0.31\pm0.03$ & $0.53\pm0.03$ \\
  $\xi_{\dm}$     & $0.63\pm0.27$ & $0.49\pm0.27$ \\
  \bottomrule
\end{tabular}
\end{table}
The $\chi^2$ value is 9.0, which for 6 degrees of freedom corresponds
to a p-value of 17\%.
The corresponding contours at the 1, 2, and $3\,\sigma$ levels are
shown in figure~\ref{aa_vs_ph_150325_rel_dm}.
The contours are determined from the changes in $\chi^2$ of the fit
relative to the optimal value. 
There is a second solution whose $\chi^2$ value is worse by $8.6$
units (figure~\ref{r_over_phsp_solution1_280425} and
table~\ref{tab:solution}). 
The second solution corresponds to a small value of $a$, which is
below the confinement scale. Such a small interaction range is
theoretically disfavored.  

The scattering phase $\delta_1$ in solution 1 is positive at the level
of about $2\,\sigma$. A positive scattering phase would correspond to
an attractive isovector $\bb$ potential. In the limit of a large
$b$-quark mass, the strong interaction potential is universal,
i.e.\ the same for the $\bb$, $\bbst$, and $\bstbst$ channels. Thus,
our results are relevant for understanding molecular states with
isospin one in all these channels, in particular $Z_b(10610)$ and
$Z_b(10650)$~\cite{Bondar:2011ev}.

\section{Conclusions}
\label{sec:conc}

We measure the $m(\bn)-m(\bp)$ mass difference to be
\begin{equation}
  \dm = (0.495\pm0.024\pm0.005)\,\mevm.
  \label{eq:dm_result}
\end{equation}
A key ingredient of this measurement is the determination of the
energy dependence of the
$\rnp=\sigma(\ee\to\bnbn)/\sigma(\ee\to\bpbm)$ cross-section ratio
that we constrain using the variation of the average $\ee$
c.m.\ energy over the data-taking periods of Belle and Belle~II. 

The result in eq.~\eqref{eq:dm_result} is significantly different from
a similar measurement by BaBar,
$\dm=0.33\pm0.05\pm0.03\,\mevm$~\cite{BaBar:2008ikz}, which currently
dominates the world average~\cite{ParticleDataGroup:2024cfk}. In
contrast to the analysis reported here, in ref.~\cite{BaBar:2008ikz}
the energy dependence of $\rnp$ was not determined experimentally;
instead, the phase-space hypothesis $\rnp=(p_{\bn}/p_{\bp})^3$ was
assumed. 
If we use the phase space hypothesis in our combined fit, we obtain a
value of $\dm$ that is consistent with the BaBar measurement. However,
the fit quality is very poor. In particular, the fit fails to describe
the $\mt$ distributions of $\bn$ and $\bp$ candidates. We exclude the
phase-space hypothesis at about the $10\,\sigma$ level. 

We find that the combined fit is sensitive to $\rnp$ in a wider energy
range than covered by the direct measurement, so we perform an
additional study to measure the dependence $\rnp(\ecm)$. Here,
$\rnp(\ecm)$ is treated as a binned distribution where the values in
each bin are determined by a modified version of our combined fit. The
resulting values are reported from the $\bb$ threshold up to
$10.59\,\gev$. We also perform a phenomenological analysis of the
$\rnp(\ecm)$ results to determine the parameters of the isovector
$\bb$ potential.  At the $2\,\sigma$ level, we find an attractive
isovector $\bb$ potential.  
Additional data closer to the $\bb$ threshold could help to improve
the precision in $\rnp$ and in the phenomenological model parameters.

\section{Acknowledgements}
This work, based on data collected using the Belle II detector, which was built and commissioned prior to March 2019,
and data collected using the Belle detector, which was operated until June 2010,
was supported by
Higher Education and Science Committee of the Republic of Armenia Grant No.~23LCG-1C011;
Australian Research Council and Research Grants
No.~DP200101792, 
No.~DP210101900, 
No.~DP210102831, 
No.~DE220100462, 
No.~LE210100098, 
and
No.~LE230100085; 
Austrian Federal Ministry of Education, Science and Research,
Austrian Science Fund (FWF) Grants
DOI:~10.55776/P34529,
DOI:~10.55776/J4731,
DOI:~10.55776/J4625,
DOI:~10.55776/M3153,
and
DOI:~10.55776/PAT1836324,
and
Horizon 2020 ERC Starting Grant No.~947006 ``InterLeptons'';
Natural Sciences and Engineering Research Council of Canada, Digital Research Alliance of Canada, and Canada Foundation for Innovation;
National Key R\&D Program of China under Contract No.~2024YFA1610503,
and
No.~2024YFA1610504
National Natural Science Foundation of China and Research Grants
No.~11575017,
No.~11761141009,
No.~11705209,
No.~11975076,
No.~12135005,
No.~12150004,
No.~12161141008,
No.~12405099,
No.~12475093,
and
No.~12175041,
and Shandong Provincial Natural Science Foundation Project~ZR2022JQ02;
the Czech Science Foundation Grant No. 22-18469S,  Regional funds of EU/MEYS: OPJAK
FORTE CZ.02.01.01/00/22\_008/0004632 
and
Charles University Grant Agency project No. 246122;
European Research Council, Seventh Framework PIEF-GA-2013-622527,
Horizon 2020 ERC-Advanced Grants No.~267104 and No.~884719,
Horizon 2020 ERC-Consolidator Grant No.~819127,
Horizon 2020 Marie Sklodowska-Curie Grant Agreement No.~700525 ``NIOBE''
and
No.~101026516,
and
Horizon 2020 Marie Sklodowska-Curie RISE project JENNIFER2 Grant Agreement No.~822070 (European grants);
L'Institut National de Physique Nucl\'{e}aire et de Physique des Particules (IN2P3) du CNRS
and
L'Agence Nationale de la Recherche (ANR) under Grant No.~ANR-23-CE31-0018 (France);
BMFTR, DFG, HGF, MPG, and AvH Foundation (Germany);
Department of Atomic Energy under Project Identification No.~RTI 4002,
Department of Science and Technology,
and
UPES SEED funding programs
No.~UPES/R\&D-SEED-INFRA/17052023/01 and
No.~UPES/R\&D-SOE/20062022/06 (India);
Israel Science Foundation Grant No.~2476/17,
U.S.-Israel Binational Science Foundation Grant No.~2016113, and
Israel Ministry of Science Grant No.~3-16543;
Istituto Nazionale di Fisica Nucleare and the Research Grants BELLE2,
and
the ICSC – Centro Nazionale di Ricerca in High Performance Computing, Big Data and Quantum Computing, funded by European Union – NextGenerationEU;
Japan Society for the Promotion of Science, Grant-in-Aid for Scientific Research Grants
No.~16H03968,
No.~16H03993,
No.~16H06492,
No.~16K05323,
No.~17H01133,
No.~17H05405,
No.~18K03621,
No.~18H03710,
No.~18H05226,
No.~19H00682, 
No.~20H05850,
No.~20H05858,
No.~22H00144,
No.~22K14056,
No.~22K21347,
No.~23H05433,
No.~26220706,
and
No.~26400255,
and
the Ministry of Education, Culture, Sports, Science, and Technology (MEXT) of Japan;  
National Research Foundation (NRF) of Korea Grants
No.~2021R1-F1A-1064008, 
No.~2022R1-A2C-1003993,
No.~2022R1-A2C-1092335,
No.~RS-2016-NR017151,
No.~RS-2018-NR031074,
No.~RS-2021-NR060129,
No.~RS-2023-00208693,
No.~RS-2024-00354342
and
No.~RS-2025-02219521,
Radiation Science Research Institute,
Foreign Large-Size Research Facility Application Supporting project,
the Global Science Experimental Data Hub Center, the Korea Institute of Science and
Technology Information (K25L2M2C3 ) 
and
KREONET/GLORIAD;
Universiti Malaya RU grant, Akademi Sains Malaysia, and Ministry of Education Malaysia;
Frontiers of Science Program Contracts
No.~FOINS-296,
No.~CB-221329,
No.~CB-236394,
No.~CB-254409,
and
No.~CB-180023, and SEP-CINVESTAV Research Grant No.~237 (Mexico);
the Polish Ministry of Science and Higher Education and the National Science Center;
the Ministry of Science and Higher Education of the Russian Federation
and
the HSE University Basic Research Program, Moscow;
University of Tabuk Research Grants
No.~S-0256-1438 and No.~S-0280-1439 (Saudi Arabia), and
Researchers Supporting Project number (RSPD2025R873), King Saud University, Riyadh,
Saudi Arabia;
Slovenian Research Agency and Research Grants
No.~J1-50010
and
No.~P1-0135;
Ikerbasque, Basque Foundation for Science,
State Agency for Research of the Spanish Ministry of Science and Innovation through Grant No. PID2022-136510NB-C33, Spain,
Agencia Estatal de Investigacion, Spain
Grant No.~RYC2020-029875-I
and
Generalitat Valenciana, Spain
Grant No.~CIDEGENT/2018/020;
the Swiss National Science Foundation;
The Knut and Alice Wallenberg Foundation (Sweden), Contracts No.~2021.0174, No.~2021.0299, and No.~2023.0315;
National Science and Technology Council,
and
Ministry of Education (Taiwan);
Thailand Center of Excellence in Physics;
TUBITAK ULAKBIM (Turkey);
National Research Foundation of Ukraine, Project No.~2020.02/0257,
and
Ministry of Education and Science of Ukraine;
the U.S. National Science Foundation and Research Grants
No.~PHY-1913789 
and
No.~PHY-2111604, 
and the U.S. Department of Energy and Research Awards
No.~DE-AC06-76RLO1830, 
No.~DE-SC0007983, 
No.~DE-SC0009824, 
No.~DE-SC0009973, 
No.~DE-SC0010007, 
No.~DE-SC0010073, 
No.~DE-SC0010118, 
No.~DE-SC0010504, 
No.~DE-SC0011784, 
No.~DE-SC0012704, 
No.~DE-SC0019230, 
No.~DE-SC0021274, 
No.~DE-SC0021616, 
No.~DE-SC0022350, 
No.~DE-SC0023470; 
and
the Vietnam Academy of Science and Technology (VAST) under Grants
No.~NVCC.05.02/25-25
and
No.~DL0000.05/26-27.

These acknowledgements are not to be interpreted as an endorsement of any statement made
by any of our institutes, funding agencies, governments, or their representatives.

We thank the SuperKEKB team for delivering high-luminosity collisions;
the KEK cryogenics group for the efficient operation of the detector solenoid magnet and IBBelle on site;
the KEK Computer Research Center for on-site computing support; the NII for SINET6 network support;
and the raw-data centers hosted by BNL, DESY, GridKa, IN2P3, INFN, 
PNNL/EMSL, 
and the University of Victoria.

\appendix

\section{\boldmath Energy dependence of $\sigma(\ee\to{}b\bar{b})$ and $\rnp$}
\label{app:xsecb_rat}

\begin{figure}[htbp]
\centering
\includegraphics[width=0.24\linewidth]{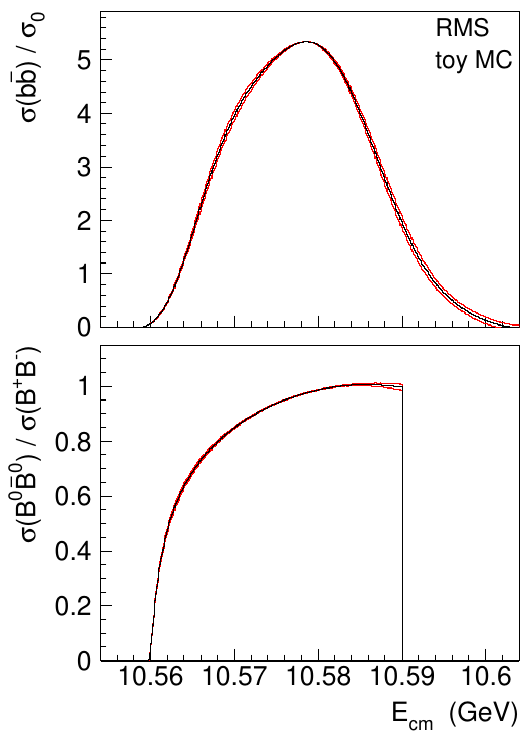}
\includegraphics[width=0.24\linewidth]{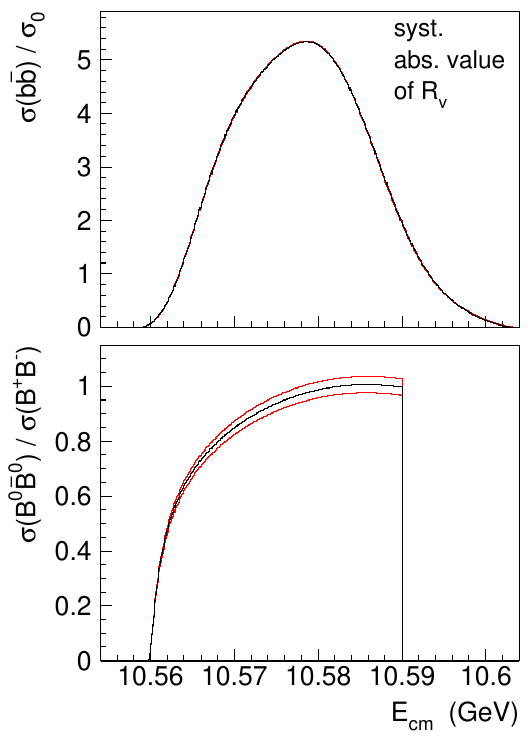}
\includegraphics[width=0.24\linewidth]{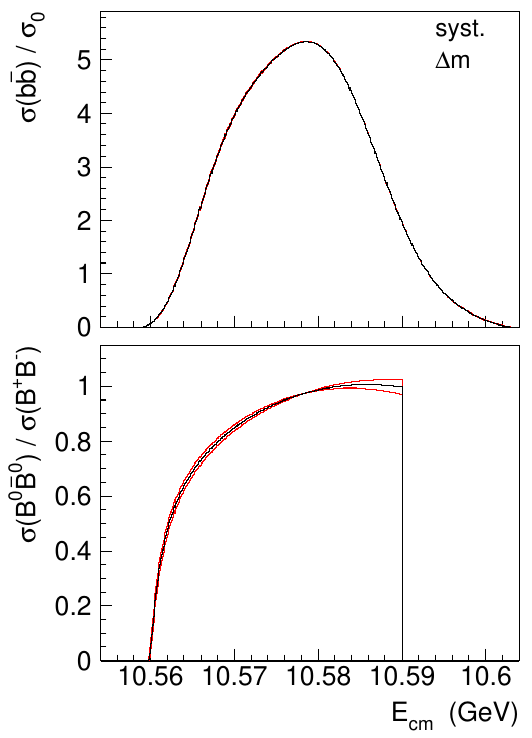}
\includegraphics[width=0.24\linewidth]{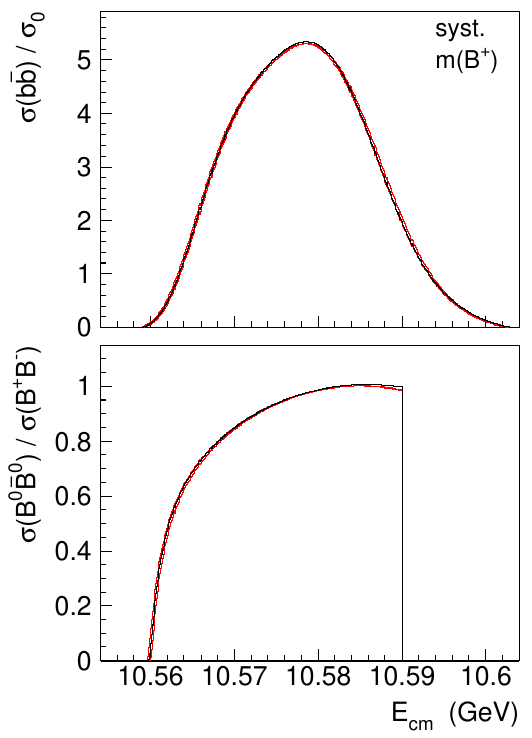}
\caption{ Energy dependence of (top)
  $[\sigma(\ee\to\bnbn)+\sigma(\ee\to\bpbm)]/\sigma_0$ and (bottom)
  $\sigma(\ee\to\bnbn)/\sigma(\ee\to\bpbm)$, where
  $\sigma(\ee\to\bnbn)$ and $\sigma(\ee\to\bpbm)$ are the
  corresponding dressed cross sections, and $\sigma_0 =
  (4\pi\alpha/3)/\ecm^2$ is the Born-level $\ee\to\uu$ cross
  section. The black curve in each plot is the result of the nominal
  fit. The red curves in the leftmost plots show the deviations
  corresponding to the RMS of the fit results for the
  pseudo-experiments. The remaining plots show deviations
  corresponding to the main sources of systematic uncertainty:
  uncertainties in $\rnpv$, $\dm$, and $m(\bp)$.} 
\label{rms_toy_150325}
\end{figure}

Figure~\ref{rms_toy_150325} shows the fitted energy dependence of
$[\sigma(\ee\to\bnbn)+\sigma(\ee\to\bpbm)]/\sigma_0$ and
$\rnp=\sigma(\ee\to\bnbn)/\sigma(\ee\to\bpbm)$, where
$\sigma(\ee\to\bnbn)$ and $\sigma(\ee\to\bpbm)$ are the corresponding
dressed cross sections, and $\sigma_0 = (4\pi\alpha/3)/\ecm^2$ is the
Born-level $\ee\to\uu$ cross section. 
In the panels of figure~\ref{rms_toy_150325}, we show the contribution
of statistical uncertainty estimated using pseudo-experiments and the
main sources of systematic uncertainty. We fit pseudo-experiments
fixing the $\dm$ value and take into account the uncertainty in $\dm$
as systematic uncertainty. The ratio $\rnp$ is shown only below
$\ecm=10.59\,\gev$, since it is not constrained by our fit above this
value.  

These plots in numerical format are given in the supplemental
material~\cite{supp_mat}. We present the values of the curves with a
step in energy of $0.1\,\mev$. To provide information on statistical
uncertainty, the material contains the fit results for 1000
pseudo-experiments.

\section{Phenomenological model}
\label{app:pheno}

The energy dependence for the dressed cross-section ratio
$\sigma(\ee\to\bnbn)/\sigma(\ee\to\bpbm)$, arising from the mass
difference $m(\bn)-m(\bp)$ and the electromagnetic interaction between
$\bp$ and $\bm$ for {\it point-like} $B$-mesons is given by the
expression 
\begin{equation}
  \rnp_0 = \left(\frac{p_{\bn}}{p_{\bp}}\right)^3\, \frac{1-\exp(-2\pi\lambda)}{2 \pi \lambda}\,
  \frac{1}{1+ \lambda^2},
  \label{eq:rnp0}
\end{equation}
where $\lambda = \alpha/(2v_c)$ is the Coulomb parameter with $v_c$
being the velocity of one charged $B$ meson in the
c.m. system~\cite{ll}.

The expression in eq.~\eqref{eq:rnp0} does not take into account two
effects: the finite size of the $B$ mesons and the scattering phase in
the isovector channel, which changes the relative phase between the
outgoing $\bnbn$ and $\bpbm$ waves. These two effects were studied in
ref.~\cite{Dubynskiy:2007xw} for $P$-wave production, where both
isospin breaking effects were treated as perturbations of the
potential at linear order. The general formula in terms of the
difference $\Delta V(r)$ of the potentials for the $\bnbn$ and $\bpbm$
pairs is 
\begin{equation}
 \rnp=  1-{ 1 \over v} \, {\rm Im}\left [ e^{2i \delta_1} \,
\int_a^\infty e^{2ipr} \, \left ( 1+ {i \over p r} \right )^2 \,
\Delta V(r) \, dr \right ]~,
\label{eq:rnp1}
\end{equation}
where $v$ and $p$ are the average $\bn$ and $\bp$ velocity and
momentum, respectively.

The lower limit $a$ in the integral in eq.(\ref{eq:rnp1}) is a
short-distance cutoff in the following sense. The difference in the
potential $\Delta V$ can be defined as long as the individual channels
$\bnbn$ and $\bpbm$ can be distinguished. However, at short distances
where the mesons overlap due to their finite spatial size, the two
channels are generally strongly mixed and thus cannot be
distinguished. Thus any difference in the potential $\Delta V$ should
vanish at short distances. This can be implemented by
introducing~\cite{Dubynskiy:2007xw} one effective cutoff distance $a$,
so that $\Delta V =0$ at $r < a$.

The potential difference due to the mass splitting and the Coulomb
interaction has the form $\Delta V = 2 \Delta m - \alpha/r$, and the
integral in eq.~(\ref{eq:rnp1}) is given as~\cite{Dubynskiy:2007xw}
\begin{eqnarray}
&&\int_{a}^\infty e^{2ipr} \,
\left ( 1+ {i \over p r} \right )^2 \, \, \Delta V(r) \, dr =  
\nonumber \\
&& - {\Delta m \over p} \left [ {2 \, \cos 2 \, p a \over p a}+ \sin 2 \, p a +
i \, \left ( {2 \sin 2 \, p a \over p a} - \cos 2 \, p a \right) \right ]   \nonumber \\
&& + \alpha \left [ {\cos 2 pa \over 2
(pa)^2}+{\sin 2 pa \over pa}-{\rm Ci}(2pa)  + i \, \left ( {\pi \over
2}-{\cos 2 pa
\over pa} + {\sin 2 pa \over 2 (pa)^2} - {\rm Si}(2pa) \right ) \right ], 
\label{eq:intv}
\end{eqnarray}
where Si$(z)$ and Ci$(z)$ are the integral sine and cosine functions.

For a quantitative comparison of the obtained expression for $\rnp$
with experiment, it is necessary to take into account the dependence
of the scattering phase $\delta_1$ on energy. The scattering phase in
a wave with angular momentum $\ell$ is found from the
formula~\cite{ll} 
\begin{equation}
\delta_1^{(\ell)} = - {\pi M \over 2} \int_0^\infty V_1(r) \left [ J_{\ell+1/2} (p r) \right ]^2 \, r \, dr~,
\label{eq:dl}
\end{equation}
where $J_\nu(z)$ is the standard Bessel function and we take $\ell=1$.
In the model~\cite{Voloshin:2018ejo}, $V_1(r)$ is assumed to take a
constant value $V_1$ in the region $r<a$ and zero outside. Therefore
eq.~\eqref{eq:dl} becomes
\begin{equation}
\delta_1 = - {\pi M \over 2} \, V_1 \int_0^a \left [ J_{3/2} (p r) \right ]^2 \, r \, dr~,
\label{eq:dl2}
\end{equation}

\end{document}